\pdfoutput=1

\documentclass[preprintnumbers,showpacs,floats,floatfix,amssymb,twocolumn,superscriptaddress,nofootinbib,prd,aps]{revtex4-1}
\usepackage{subfigure}
\usepackage{graphicx}

\newcommand{\be}{\begin{equation}}
\newcommand{\ee}{\end{equation}}
\newcommand{\beq}{\begin{eqnarray}}
\newcommand{\eeq}{\end{eqnarray}}
\newcommand{\eqref}[1]{(\ref{#1})}

\makeatletter

\newcommand{\arXiv}[2][]{\href{http://arxiv.org/abs/#2}{\texttt{arXiv:#2\@ifempty{#1}{}{ [#1]}}}}
\makeatother

\begin{document}
\title{Two-point correlators revisited:\\ Fast and slow scales in multifield models of inflation}

\author{Jos\'{e} T. G\'{a}lvez Ghersi}%
\email{joseg@sfu.ca}
\affiliation{Department of Physics, Simon Fraser University,\\
8888 University Drive, Burnaby, British Columbia V5A 1S6, Canada}
\date{\today}

\author{Andrei V. Frolov}%
\email{frolov@sfu.ca}
\affiliation{Department of Physics, Simon Fraser University,\\
8888 University Drive, Burnaby, British Columbia V5A 1S6, Canada}

\begin{abstract}
We study the structure of two-point correlators of the inflationary field fluctuations in order to improve the accuracy and efficiency of the existing methods to calculate primordial spectra. We present a description motivated by the separation of the fast and slow evolving components of the spectrum which is based on Cholesky decomposition of the field correlator matrix. Our purpose is to rewrite all the relevant equations of motion in terms of slowly varying quantities. This is important in order to consider the contribution from high-frequency modes to the spectrum without affecting computational performance. The slow-roll approximation is not required to reproduce the main distinctive features in the power spectrum for each specific model of inflation.       
\end{abstract}
\maketitle
\section{Introduction}
\begingroup
The Inflationary paradigm has become an important piece in our understanding of the Early Universe. It has been designed and constantly improved since its very first appearance in Ref.~\cite{Guth:1980zm} wherein it solves the main issues of standard Big Bang Cosmology. In the last three decades, we witnessed the emergence of a plethora of models, which not only aim to provide expansion for a sufficient number of e-folds, but also expect to produce a power spectrum of fluctuations consistent with the actual structure of the universe.

Direct computation of the power spectrum from equations of motion can be time consuming, especially when resolving specific features of each model in a wide range of energy scales. These problems will be aggravated for multifield models of inflation such as the proposals in Refs.~\cite{Greene:1997fu, Tsujikawa:2002qx}.

The primordial power spectrum is valuable in finding sensible ranges of validity for the parameters of any proposed Inflationary model. In the case of models with multiple fields, it encodes vital information about the power transfer between different components. The viability of a model is usually tested when the spectrum is loaded as an input in any of the existing schemes based in Boltzmann transport equations (such as CAMB in Ref.~\cite{Challinor:2005jy} or CLASS in Ref.~\cite{Tram:2013ima}) and compared with data. Being aware of the existing difficulties to design a spectral code, we suggest an approach intended to isolate all the high frequency terms and only use the slowly evolving quantities relevant for calculating the spectrum. These degrees of freedom remain ``frozen''  outside the horizon, which implies the use of large time steps in any numerical evolution scheme.

A crucial part of separating fast and slow degrees of freedom relies on focusing on the spectrum, which contains information about all the field correlations. This spectrum is generated by the symmetric product between two field multiplets, forming a correlation matrix. Field dynamics reveals the approximate time-translational invariance of each multiplet component on subhorizon scales. This symmetry must be exploited to define a well-posed Cauchy problem each time vacuum correlations are defined as initial conditions. When the cases of anticorrelation are excluded, positive definite correlation matrices are suitable for Cholesky decomposition into two unique triangular factors. Anticorrelations break the uniqueness of this factorization. This decomposition has been the preferred tool of statisticians to generate correlated samples from any set of unit variance random vectors. Amplitudes of the modes and power of the cross correlations are carried in these Cholesky factors, which act on a rotating basis of solutions, just as in the Schrodinger picture in Quantum Mechanics. The fast rotation of this basis is separated from the slow evolution of amplitudes. We implement a dynamical Cholesky decomposition motivated by the separation of wave solutions into phases and amplitudes. This separation has been explored in single field solutions, (see Refs.~\cite{Brooker:2015iya, Bardeen:1985tr} for more details) with exact results in the case of massless perturbations.   
  
The plan for this paper is as follows: in section \ref{sec:review}, we will review the notions of perturbation theory for inflationary models, along with the equations of motion for the background fields. In section \ref{sec:correlators}, we describe the field decomposition technique used in order to separate the fast oscillating phases from the amplitudes. The latter are required to calculate the spectrum. In section \ref{sec:initial}, we discuss the background dynamics and the use of initial conditions based on asymptotic vacuum solutions. Additionally, we introduce a scheme for injecting modes in the system. To conclude, we present our results and discussions.
\endgroup

\vfill

\section{Perturbation theory: a lightning review}\label{sec:review}
\begingroup
\exhyphenpenalty 10000
For the purposes of this work, we only consider Lagrangian densities with canonical kinetic terms. We will later extend our treatment in Appendix \ref{App:AppendixA} for the case of a generic curved field space developed in Ref.~\cite{Sasaki:1995aw}.  The action $S$ needed to describe a generic model of inflation with multiple fields coupled reads,
\be
S=\int d^4x\sqrt{-g}\left(\frac{M_{\text{Pl}}^2}{2}R-\frac{1}{2}\partial_\mu\phi_A\partial^\mu\phi_A-V(\phi_A)\right),\label{action}
\ee
where all the fields become coordinates of a generically non-flat field space. We pick spatially flat coordinates just as described in Ref.~\cite{Sasaki:1995aw} 
\be
ds^2=-(1+2A)dt^2-2a^2\partial_iB dx^i dt+a^2\delta_{ij}dx^idx^j,
\label{spflat}
\ee
with the $(-+++)$ signature. Now, as is usual in this perturbative approach, we decompose each component of the $\mathcal{N}$ field multiplet $\phi_A$ into a spatially homogeneous background field and its fluctuations
\be
\phi_A(x^\mu)=\phi^0_A(t)+\tilde{\Phi}_A(x^\mu).\nonumber
\ee 
At the background level, the field equations given by (\ref{action}) read
\be
\ddot{\phi}^0_A+3H\dot{\phi}^0_A+\frac{\partial V}{\partial\phi_A}=0. \label{back}
\ee
The expansion history is obtained from the first Friedmann equation
\be
3M_{\text{Pl}}^2H^2=\frac{1}{2}\dot{\phi}^0_A\dot{\phi}^0_A+V(\phi^0_A).\label{expansion}
\ee
where we define $\dot{\phi}\equiv d\phi/dt$. These equations describe all the quantities needed from the homogeneous limit -- such as the scale factor and the masses --  and hence, these become a part of the overall evolution scheme.

In order to find the spectrum, quantum fluctuations are normally expressed as Fourier modes $\Phi_A(k,t)$ evaluated on a constant time surface, where we are able to set the scale as a constant $a$ without loss of generality. These modes are contained in a second order expansion of (\ref{action})
\be
S_\Phi(k)=\frac{1}{2}\int a^3\, dt\left[\dot{\Phi}_A\dot{\Phi}_A-\Phi_A\left(\frac{k^2}{a^2}\delta_{AB}+M^2_{AB}\right)\Phi_B\right],\label{pertact}
\ee
where $M^2_{AB}\equiv\partial^2V/\partial\phi_A\partial\phi_B-\left(a^3\dot{\phi}_A\dot{\phi}_B/H\right)^./a^3M_{\text{Pl}}^2$ is the well-known ``mass-squared matrix''  modified by the coordinate choice made on (\ref{spflat}). All the equations of motion just follow from variational principle. Initial conditions will be consistent with the high-frequency behaviour $k/a\gg M^2_{AB}$, where the solutions for $\Phi_A$ are treated as in Minkowski spacetime.   
\endgroup

\section{Two-point correlators revisited}\label{sec:correlators}
It is important to compute the spectrum of linearized field perturbations at the end of inflation since it is directly related to the spectrum of primordial curvature. An interesting procedure for this calculation is described in Ref.~\cite{Price:2014xpa}, where the authors consider the evolved components of a complex ``mode matrix'', which follows the same equations of motion obtained from (\ref{pertact}). More recently, the Hamiltonian evolution of the field correlators $\langle\Phi_A,\Phi_B\rangle$ was considered in Ref.~\cite{Dias:2016rjq} with the purpose of separating fast and slow evolution scales. In these perspective, more than one time scale is still necessary to resolve the mode correlations.

Here we present a different approach: a dynamical decomposition method which suppresses the fastest oscillation scales in the equations of motion of the correlation modes regardless of the mode frequency. Hence, the evolution of the correlation modes is more efficient, particularly for deep sub-horizon scales where integration costs scale linearly with number of e-folds in our method, as opposed to exponential scaling in other methods. To achieve that, we suggest the following extended form of the field 
\be
\Phi_A=L_{AB}\,\hat{\chi}_B\label{ansatz}
\ee        
as an effective way to separate a real matrix of amplitudes $(L_{AB})$ from phases $(\hat{\chi}_B)$ on each mode. We initially assume $\mathcal{N}\times \mathcal{N}$ independent degrees of freedom contained by $L_{AB}$. Field fluctuations in de Sitter spacetime before crossing the horizon only add ``damping'' to the standard massive wave solutions in empty space. For that reason, it is possible to consider the vector $\hat{\chi}_B$ as a set of evolved phases.
By inserting the ansatz (\ref{ansatz}) in (\ref{pertact}), we get
\beq
S_\Phi&=&\frac{1}{2}\int a^3\, dt\delta_{AB}\bigg[L_{AC}L_{BD}\:\dot{\hat{\chi}}_C\:\dot{\hat{\chi}}_D\nonumber\\
&+&2\dot{L}_{BD}L_{AC}\:\hat{\chi}_D\:\dot{\hat{\chi}}_C+\dot{L}_{AC}\dot{L}_{BD}\:\hat{\chi}_C\:\hat{\chi}_D\bigg]\nonumber\\
&-&L_{AC}\left(\frac{k^2}{a^2}\delta_{AB}+M^2_{AB}\right)L_{BD}\:\hat{\chi}_C\:\hat{\chi}_D\label{actchi}
\eeq   
and the equations of motion for $\hat{\chi}_E$ are
\beq
&\ddot{\hat{\chi}}_E+\left(3H\delta_{EB}+2\left(L^{-1}\right)_{ED}\dot{L}_{DB}\right)\dot{\hat{\chi}}_B+\nonumber\\
&\bigg[\left(L^{-1}\right)_{ED}\ddot{L}_{DB}+3H\left(L^{-1}\right)_{ED}\dot{L}_{DB}+\nonumber\\
&\left(L^{-1}\right)_{ED}\left[\frac{k^2}{a^2}\delta_{DC}+M^2_{DC}\right]L_{CB}\bigg]\hat{\chi}_B=0\label{eqmov}
\eeq
which turn out to be minimally different from the case of Coriolis' equation written in a rotating non-inertial frame. The last term corresponds to the effective rotation frequency. Calculating the conjugate momentum of $\hat{\chi}_B$ from this action, we find $\hat{P}_B = a^3\left(L_{AB}L_{AD}\:\dot{\hat{\chi}}_D+\dot{L}_{AD}L_{AB}\:\hat{\chi}_D\right)$, hence the canonical commutators $\left[\hat{\chi}_A,\hat{P}_B\right]=i\delta_{AB}$ now imply
\be
\left[\hat{\chi}_A,\dot{\hat{\chi}}_B\right]=\frac{i}{a^3}(L^T)^{-1}_{BC}(L^{-1})_{CA}.
\label{comm}
\ee    
 
The dependence on $a^{-3}$ is responsible for the suppressed commutators during inflation. This is consistent with the decoherence conditions stated in Ref.~\cite{Polarski:1995jg}. Transitioning into classical states has dynamical consequences: $\hat{\chi}_A$ can be treated as in Ref.~\cite{Bardeen:1985tr}, Gaussian random variables in which the central limit theorem is fully applicable and a complex phase is convenient in evolving two copies of the same field under different initial conditions. As a result, complexification is no longer a requirement. In addition to this, our prescription in (\ref{ansatz}) allows us to rewrite the two-point correlators as
\be
\langle\Phi_A,\Phi_B\rangle=L_{AC}L^T_{DB}\langle\hat{\chi}_C,\hat{\chi}_D\rangle.
\label{corr}
\ee
 
A similar separation is performed in Ref.~\cite{Seery:2012vj}. It is inspired by geometrical optics and considers the evolution of the correlators by describing the elements -- i.e. rotations, expansions and shears-- of the unitary evolution operator acting on each mode. However, our approach is more concerned on the dynamical constraints required to hold a specific gauge choice after the separation in (\ref{ansatz}).

Following the Central Limit Theorem, we can demand independence between different phases (i.e. a normalized covariance matrix) at all times. This is analog to using ``cosine'' or ``sine'' waves as an orthonormal basis for a simple harmonic oscillator with the purpose of maximizing \textit{amplitude} and \textit{velocity} as initial conditions for the oscillating system. This independence condition is our gauge choice
\be
\langle\hat{\chi}_C,\hat{\chi}_D\rangle=\delta_{CD},\label{firstgauge}
\ee
and it must be held at all times. The expression for the correlator in (\ref{corr}) reduces to
\be
\langle\Phi_A,\Phi_B\rangle=L_{AC}L^T_{CB}.\label{gaugedcorr}
\ee
Because phases are fast oscillating degrees of freedom and can cancel out, these are not required to calculate two-point correlators. All information of the correlations is carried by $L_{BC}$, the slowly varying ``square root'' of the correlation matrix. We look for an effective set of equations of motion which mainly depend on amplitudes instead of phases. Therefore, we must consider that the evolution of $\hat{\chi}_A$ follows the second order differential equations in (\ref{eqmov}). Hence, time translational invariance requires two derivatives of (\ref{firstgauge}) as additional constraints for the system at an arbitrary time slice 
\beq
&\text{Sym}^{CD}_{AB} \langle\dot{\hat{\chi}}_{C},\hat{\chi}_{D}\rangle=0,\label{secondgauge}\\
&\text{Sym}^{CD}_{AB} \left[\langle\ddot{\hat{\chi}}_{C},\hat{\chi}_{D}\rangle+\langle\dot{\hat{\chi}}_{C},\dot{\hat{\chi}}_{D}\rangle\right]=0,\label{thirdgauge}
\eeq    
where we define $(\text{Anti})\text{Sym}^{CD}_{AB}\equiv \delta^C_A\delta^D_B\mp\delta^C_B\delta^D_A$ and its antisymmetric analog as (anti-)symmetrizing multilinear operators. Equations (\ref{firstgauge}) and (\ref{secondgauge}) are just constrained by initial conditions. With (\ref{eqmov}), we can rewrite (\ref{thirdgauge}) as    
\beq
&\text{Sym}^{CD}_{AB}\bigg[\left(L^{-1}\right)_{CE}\ddot{L}_{ED}+3H\left(L^{-1}\right)_{CE}\dot{L}_{ED}+\nonumber\\&\left(L^{-1}\right)_{CE}\left[\frac{k^2}{a^2}\delta_{EF}+M^2_{EF}\right]L_{FD}-\langle\dot{\hat{\chi}}_C,\dot{\hat{\chi}}_D\rangle\nonumber\\
&+(L^{-1})_{CE}\dot{L}_{EF}\langle\dot{\hat{\chi}}_F,\hat{\chi}_{D}\rangle\bigg]=0\label{newgauge}.
\eeq
In all the terms, symmetrization in both free indices indicates the existence of $\mathcal{N}(\mathcal{N}-1)/2$ repeated degrees of freedom in the system. Only when the cases of anticorrelation are discarded, positive definiteness of the correlation matrix is useful to eliminate the redundancy by picking $L_{AB}$ to be Cholesky matrices. These do not have more than $\mathcal{N}(\mathcal{N}+1)/2$ non-zero real elements in the lower triangular corner of the matrix. Our choice of Cholesky matrices is convenient for a variety of reasons, especially since matrix inversions and other operations are computationally inexpensive and easily scalable for the case of models with a large number of fields.

\begin{figure*}
\begin{minipage}{\textwidth}
\includegraphics[width=0.71\textwidth]{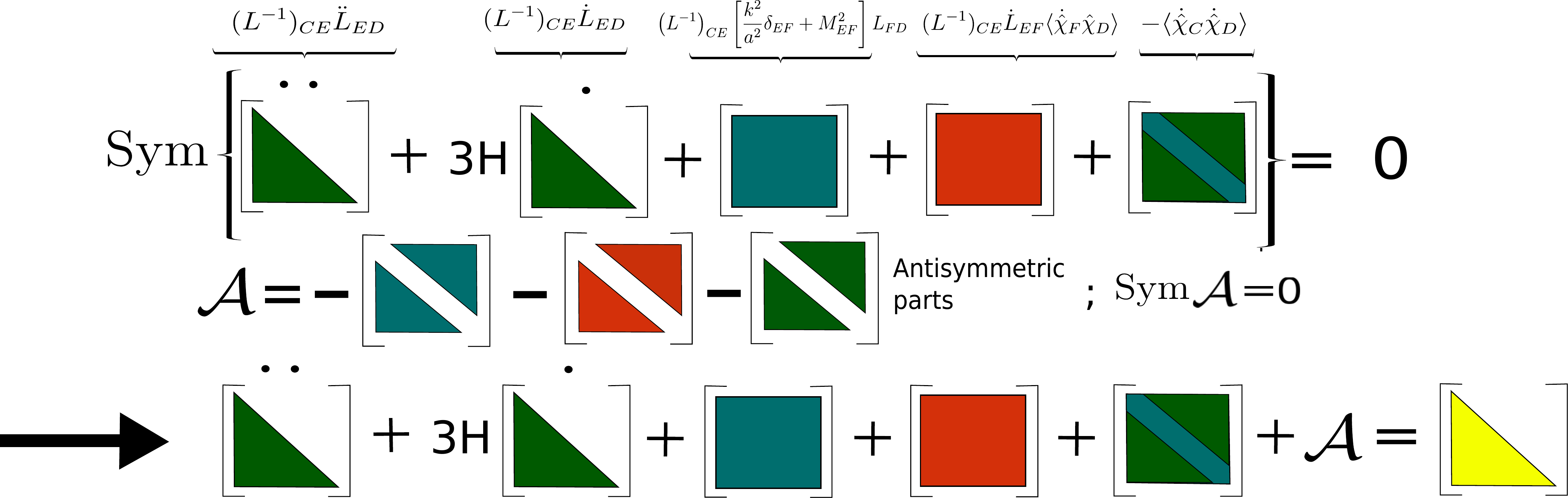}
\caption{Schematic representation of (\ref{newgauge}). Fixing the shape of Cholesky amplitudes by adding the antisymmetric matrix $\mathcal{A}_{CD}$. This matrix is an antisymmetrized collection of all components precluding out of the lower triangular form. After rearranging terms in the final sum, the expression in (\ref{gaugefinal}) keeps its Cholesky shape at all times.}
\label{scheme}
\end{minipage}
\end{figure*}

Moreover, the largest term in the squared brackets of equation (\ref{newgauge}) has a gauge symmetry due to the overall symmetrization acting on it. Such a gauge freedom can be exploited to fix the shape of the Cholesky representation at all times. Only the first two terms in this expression preserve the shape of a Cholesky matrix. As the system evolves, all the other terms will migrate degrees of freedom out of the lower triangular form of $L$. Consequently, we need to add an antisymmetric matrix $\mathcal{A}_{CD}$, which will be canceled out when symmetrized. In Fig.~\ref{scheme}, we represent the way in which the Cholesky representation is preserved.

Gauge fixing condition emerges by supressing the upper diagonal terms $(D>C)$ 
\beq
&\mathcal{A}_{CD}=-\left(L^{-1}\right)_{CE}M^2_{EF}L_{FD}-\\\nonumber
&(L^{-1})_{CE}\dot{L}_{EF}\langle\dot{\hat{\chi}}_F,\hat{\chi}_{D}\rangle+\langle\dot{\hat{\chi}}_C,\dot{\hat{\chi}}_D\rangle.
\label{gaugefixing}
\eeq 
Thus, Cholesky matrices evolve in agreement with
\beq
&\bigg[\ddot{L}_{AB}+3H\dot{L}_{AB}+\left[\frac{k^2}{a^2}\delta_{AC}+M^2_{AC}\right]L_{CB}\bigg]\nonumber\\
&+L_{AC}\mathcal{A}_{CB}+\dot{L}_{AC}\langle\dot{\hat{\chi}}_C,\hat{\chi}_B\rangle-L_{AC}\langle\dot{\hat{\chi}}_C,\dot{\hat{\chi}}_B\rangle=0\label{gaugefinal}.
\eeq
To find appropriate expressions for the expected phase correlators $\langle\dot{\hat{\chi}}_A\dot{\hat{\chi}}_B\rangle$ and $\langle\dot{\hat{\chi}}_A\hat{\chi}_B\rangle$, we will write the derivatives for each one of the unknown correlators
\beq
&\frac{d}{dt}\langle\dot{\hat{\chi}}_A,\dot{\hat{\chi}}_B\rangle=\text{Sym}^{CD}_{AB}\langle\ddot{\hat{\chi}}_{C},\dot{\hat{\chi}}_{D}\rangle\label{dotdot}\\
&\frac{d}{dt}\langle\dot{\hat{\chi}}_A,\hat{\chi}_B\rangle=\langle\ddot{\hat{\chi}}_{A},\hat{\chi}_{B}\rangle+\langle\dot{\hat{\chi}}_{A},\dot{\hat{\chi}}_{B}\rangle\nonumber.
\eeq
The last expression is equivalent to
\beq
&\frac{d}{dt}\langle\dot{\hat{\chi}}_A,\hat{\chi}_B\rangle=\frac{1}{2}\text{Sym}^{CD}_{AB}\left[\langle\ddot{\hat{\chi}}_{C},\hat{\chi}_{D}\rangle+\langle\dot{\hat{\chi}}_{C},\dot{\hat{\chi}}_{D}\rangle\right]\nonumber\\
&+\frac{1}{2}\text{Antisym}^{CD}_{AB}\left[\langle\ddot{\hat{\chi}}_{C},\hat{\chi}_{D}\rangle+\langle\dot{\hat{\chi}}_{C},\dot{\hat{\chi}}_{D}\rangle\right]\nonumber\\
&=\frac{1}{2}\text{Antisym}^{CD}_{AB}\langle\ddot{\hat{\chi}}_{C},\hat{\chi}_{D}\rangle
\label{dotchi}
\eeq
Equation (\ref{dotchi}) conserves the antisymmetric properties of $\langle\dot{\hat{\chi}}_A,\hat{\chi}_B\rangle$ found in (\ref{secondgauge}), whose behavior seems to match the rotation generators in a real vector space. Using (\ref{eqmov}) and (\ref{gaugefinal}), we can write equations of motion for the missing two-point correlators
\beq
&\frac{d}{dt}\langle\dot{\hat{\chi}}_A,\dot{\hat{\chi}}_B\rangle=\text{Sym}^{CD}_{AB}\bigg[-3H\langle\dot{\hat{\chi}}_{C},\dot{\hat{\chi}}_{D}\rangle\label{eqmovdotdot}\\
&-2\left(L^{-1}\right)_{CE}\dot{L}_{EF}\langle\dot{\hat{\chi}}_{F},\dot{\hat{\chi}}_{D}\rangle-\mathcal{A}_{CE}\langle\dot{\hat{\chi}}_E,\hat{\chi}_{D}\rangle\nonumber\\
&+\left(L^{-1}\right)_{CE}\dot{L}_{EF}\langle\dot{\hat{\chi}}_{F},\hat{\chi}_{G}\rangle\langle\hat{\chi}_{G},\dot{\hat{\chi}}_{D}\rangle+\langle\dot{\hat{\chi}}_{C},\dot{\hat{\chi}}_{F}\rangle\langle\dot{\hat{\chi}}_{F},\hat{\chi}_{D}\rangle\bigg],\nonumber
\eeq
and also
\beq
&\frac{d}{dt}\langle\dot{\hat{\chi}}_A,\hat{\chi}_B\rangle = \mathcal{A}_{AB}-3H\langle\dot{\hat{\chi}}_A,\hat{\chi}_B\rangle-\nonumber\\
&\frac{1}{2}\text{Antisym}^{CD}_{AB}\bigg[\left(L^{-1}\right)_{CE}\dot{L}_{EF}\langle\dot{\hat{\chi}}_F,\hat{\chi}_{D}\rangle\bigg]\label{eqmovdotchi}
\eeq
At first glance, the symmetrizer doubles the first coefficients in (\ref{eqmovdotdot}). After comparing (\ref{eqmovdotdot}) with (\ref{eqmovdotchi}) we notice that $\langle\dot{\hat{\chi}}_A,\dot{\hat{\chi}}_B\rangle \sim \langle\dot{\hat{\chi}}_A,\hat{\chi}_B\rangle^2$, which is consistent with our analogy with the rotation generators in a real vector space. However, a careful review of all the terms in both expressions reveals that this is not an accurate statement. Henceforth, both correlators are considered as separate elements in the evolving system. The expressions found in (\ref{gaugefinal}), (\ref{eqmovdotdot}) and (\ref{eqmovdotchi}) constitute the system of differential equations required to evolve the ``square root'' of the two-point correlators. 
\section{Initial conditions and spectral evolution scheme}\label{sec:initial}
It is reasonable to ask about the proper set of values required to initialize the equations of motion as described in (\ref{back}) and (\ref{gaugefinal}). To do so, we first consider the evolution of the background fields.  

\begin{figure*}
\begin{center}
\subfigure[]{\includegraphics[scale=0.3]{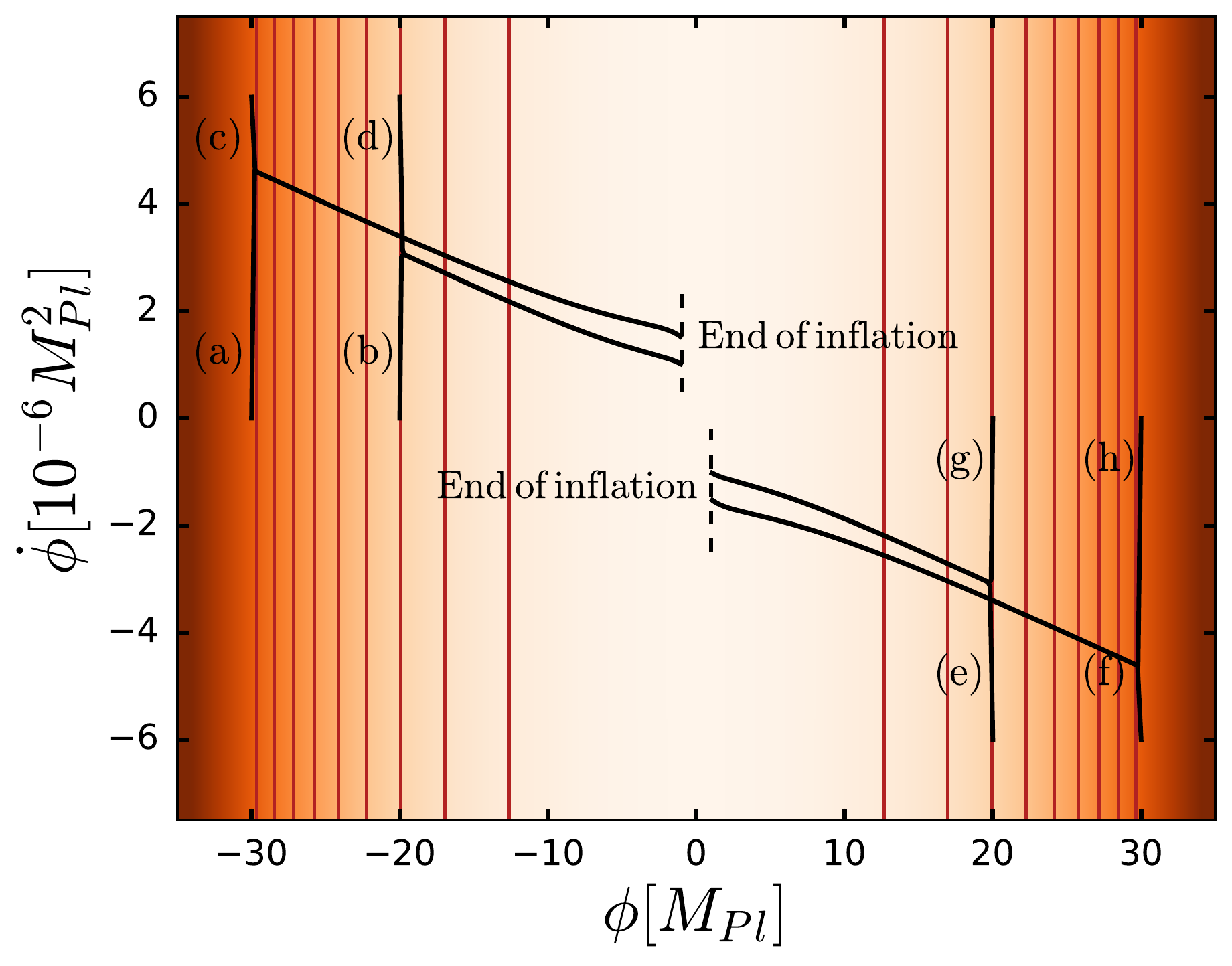}\label{proj1}}
\subfigure[]{\includegraphics[scale=0.3]{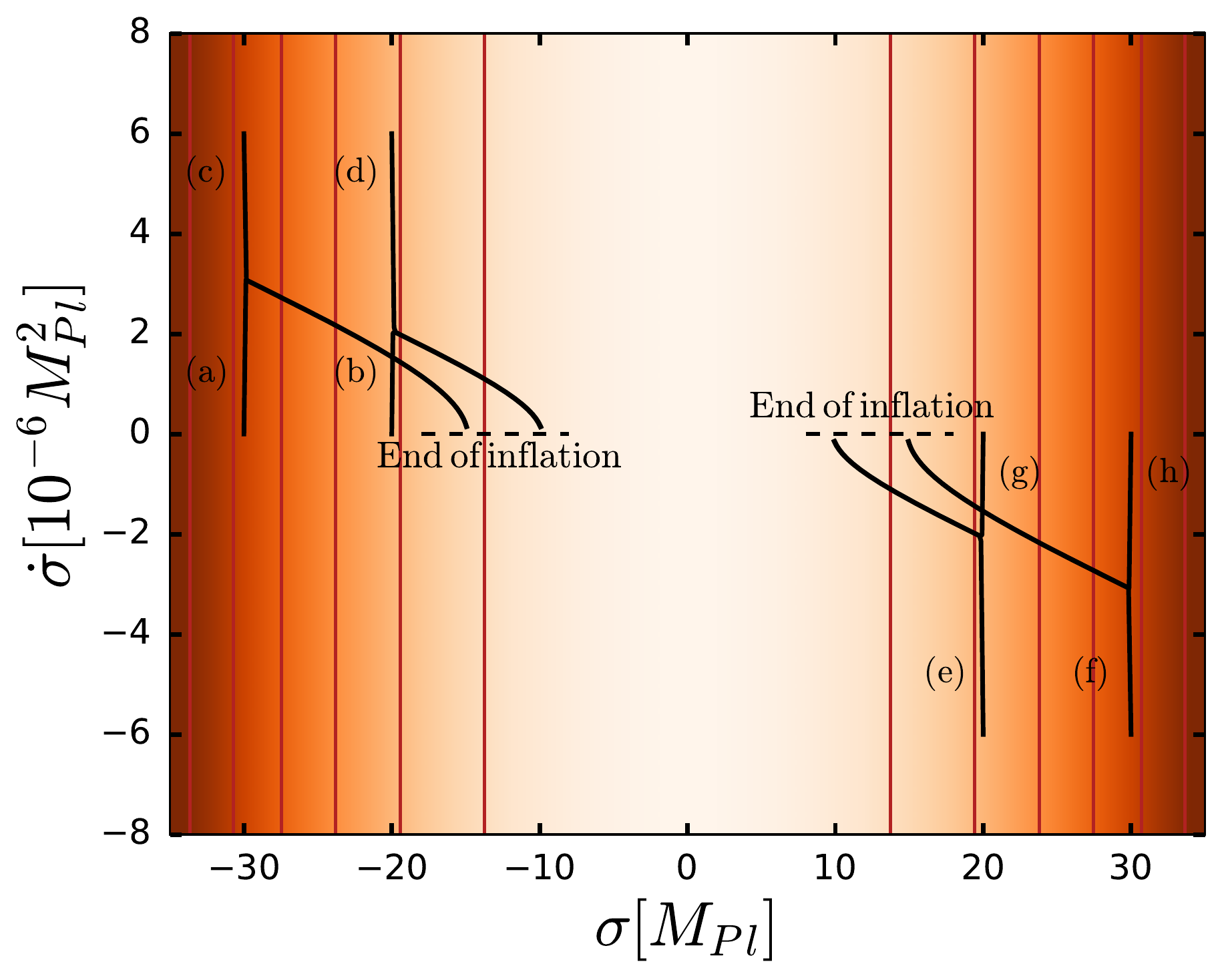}\label{proj2}}
\subfigure[]{\includegraphics[scale=0.3]{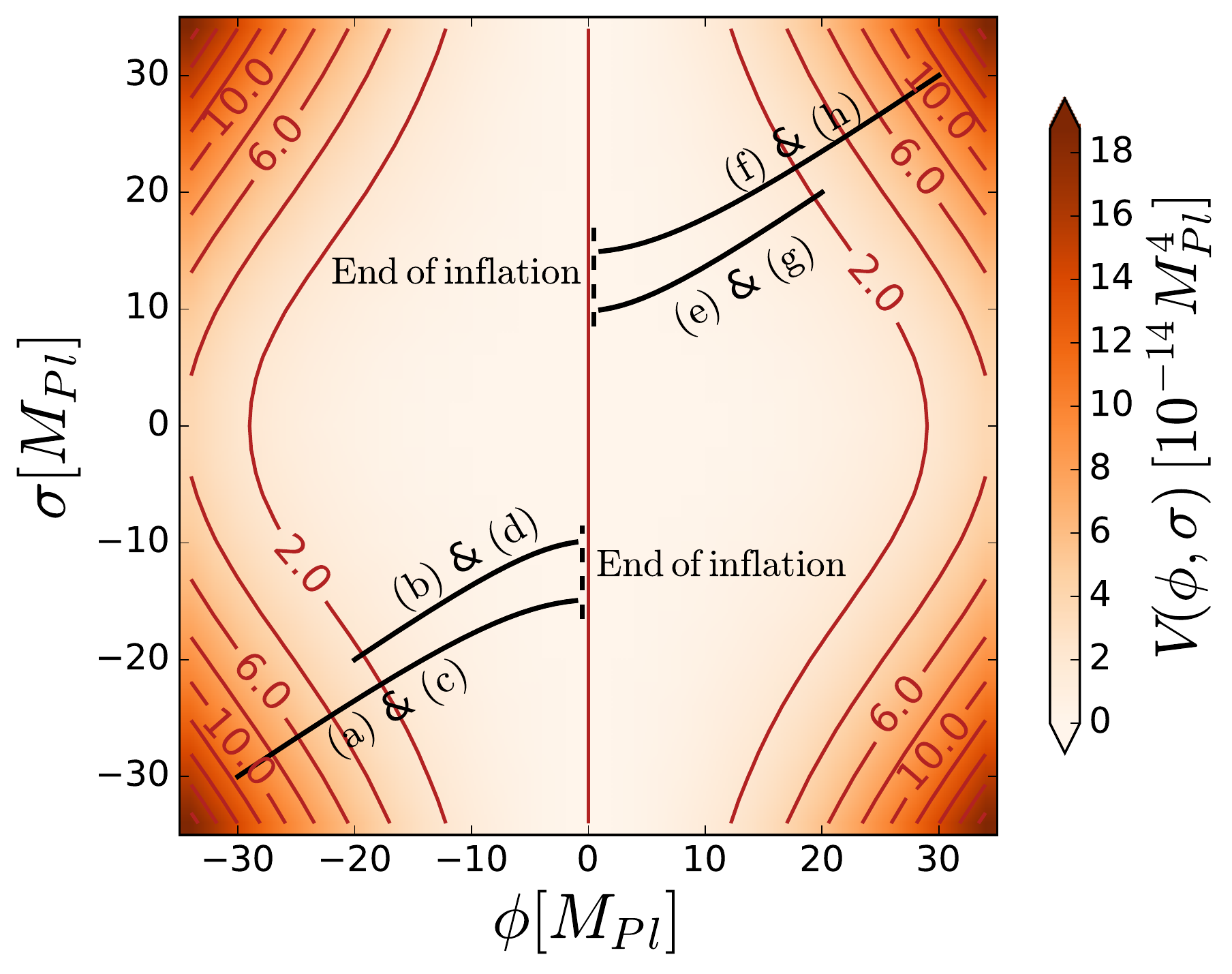}\label{proj3}}
\end{center}
\caption{Evolution of eight different background trajectories labeled from (a) to (h) and represented in three transversal sections of the phase space: $(\dot{\phi},\phi)$ in Fig.~\ref{proj1}, $(\dot{\sigma},\sigma)$ in Fig.~\ref{proj2} and $(\phi,\sigma)$ in Fig.~\ref{proj3}. The projection $(\phi,\sigma)$ is embedded on the field potential. Here we observe the overlap of trajectories with the same initial values of $\phi{(t_0)}$ and $\sigma{(t_0)}$. Equipotential curves of (\ref{lp4}) are plotted in the three projections as a reference for the energy scale at different field configurations.}
\label{fig:phasespace}
\end{figure*}

Alternative choices of a time scale (such as the number of e-folds $N\equiv \log a$) can only be implemented without affecting the background evolution before the end of inflation. The application of such scales does not affect the overall performance of any evolution scheme. Throughout the rest of the paper, we choose
\be
V(\phi,\sigma)=\frac{\lambda}{4}\phi^4+\frac{g}{2}\phi^2\sigma^2,
\label{lp4}
\ee
as an example of a multifield non-linear potential. Having a system of highly coupled second order differential equations, we must investigate the behaviour of first and second derivatives for both fields since those provide initial conditions for the background. Thus, we can observe in Fig.~\ref{fig:phasespace} three phase space projections of the background field trajectories,  
where $\lambda=g/2=10^{-14}$  is the typical numerical value assigned to the coupling constant. From Fig.~\ref{fig:phasespace}, we notice that $\dot{\phi}(t_0)\approx \dot{\sigma}(t_0)\approx 0.0$ are suitable choices for the derivatives of the slowly rolling background field. Other values will not modify the convergence time substantially. The lapse of time the system takes to fall into the inflationary attractor is very small. However, we always have the option of getting closer to the attractor in order to reduce the convergence time of the complete system of equations, which includes fluctuations. Hence, we are left with two numbers we have to fix in order to achieve enough expansion in the model.

\begin{figure}
\begin{center}
\includegraphics[width=0.48\textwidth]{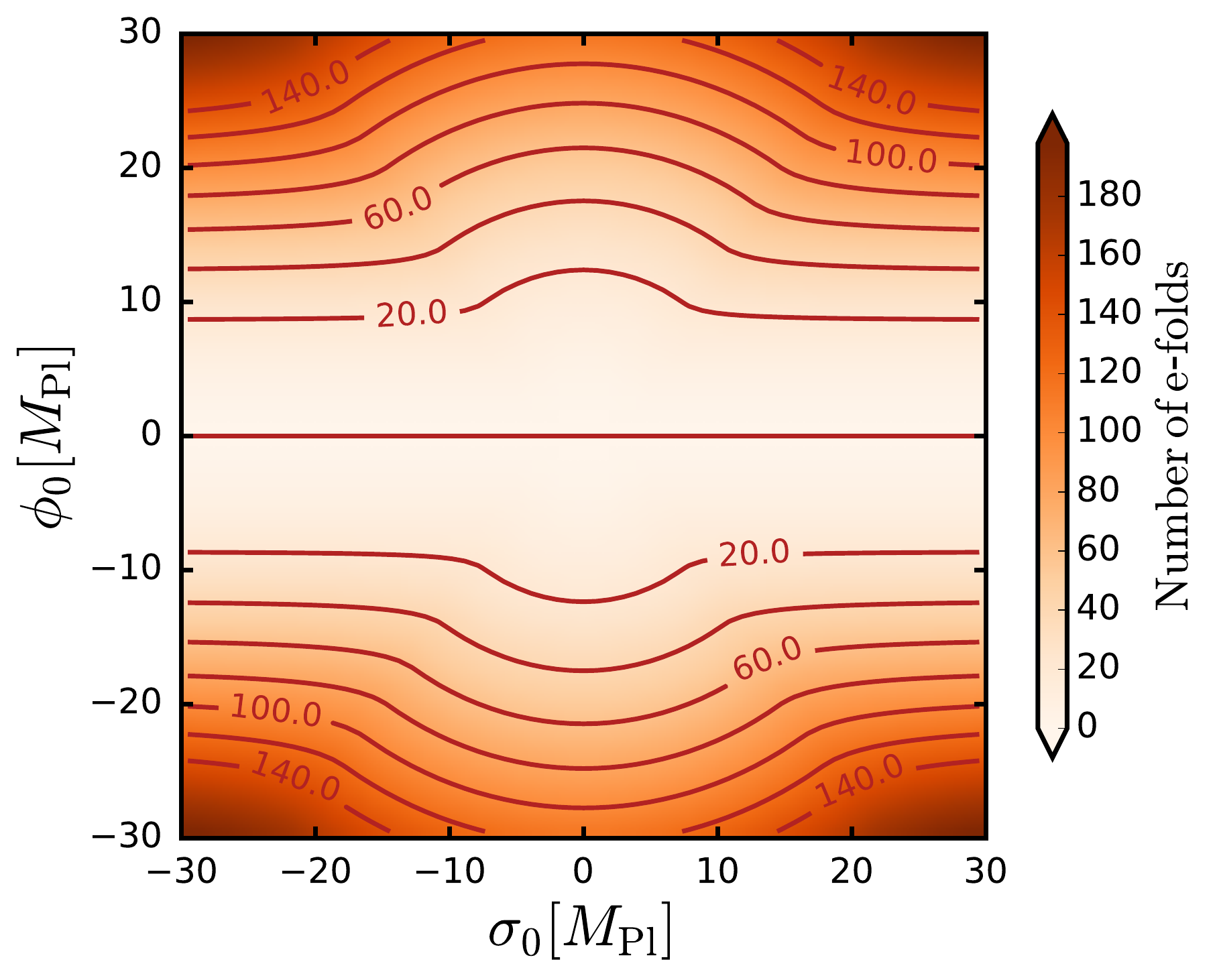}
\caption{Projected surface of initial conditions. The contours plotted highlight the regions (similar to equipotential curves) which would generate 20, 60, 100 and 140 e-folds of inflation.}
\label{figureII}
\end{center}
\end{figure}

\begin{figure}
\begin{center}
\includegraphics[width=0.45\textwidth]{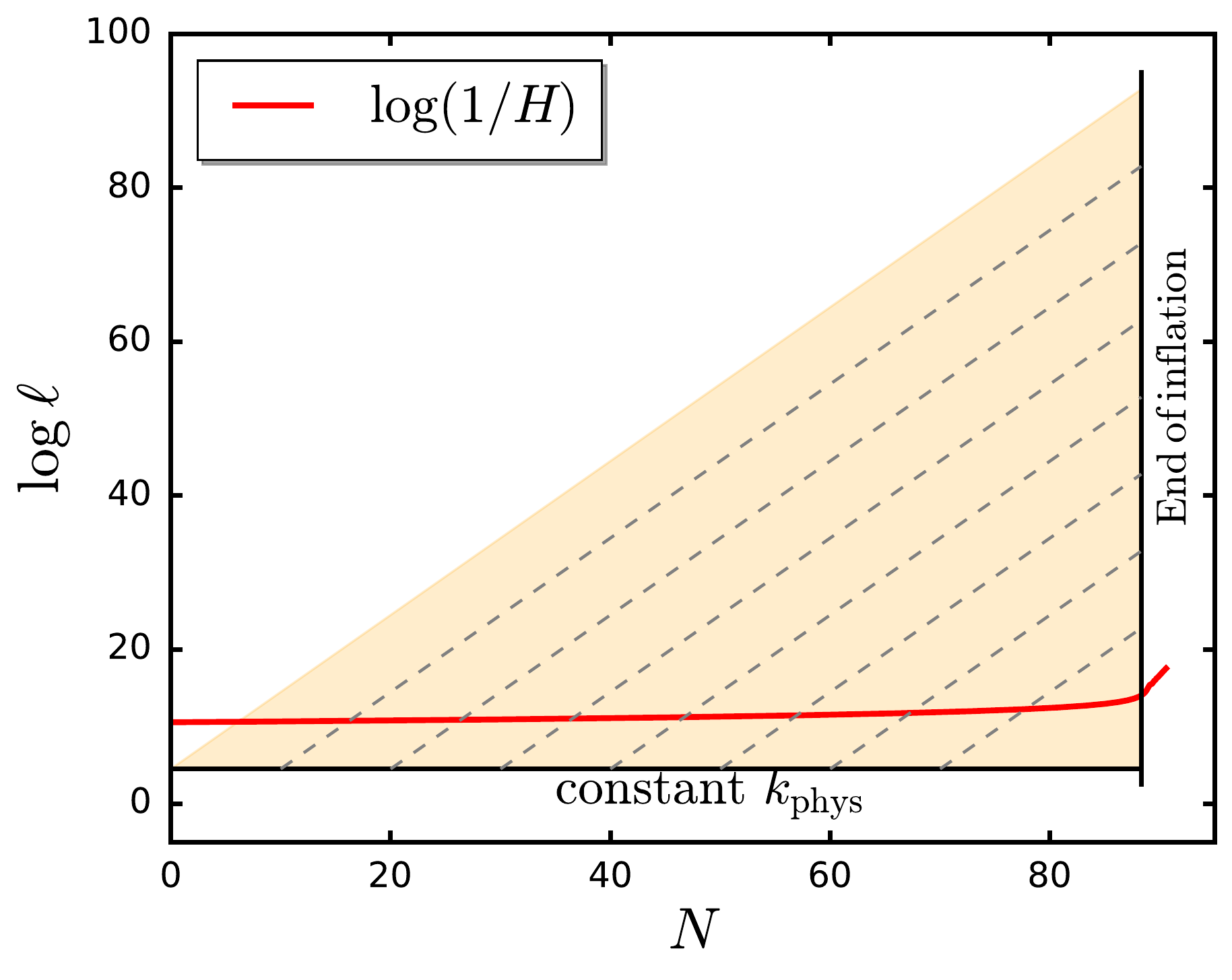}
\caption{Mode injection scheme. Cholesky modes injected every 10 e-folds from a surface of constant physical wavelength and evolved until a fixed time ``screen'' at the end of inflation.  Initial conditions allow up to 88 e-folds of inflation. $\ell$ is the physical length scale in the system.}
\label{fig:figureIV}
\end{center}
\end{figure}

In Fig.~\ref{figureII}, we generated a ``map'' of initial values for $\phi$ and $\sigma$, including the number of expanded e-folds.
Once we choose a set of initial field coordinates, we can ensure the conditions necessary to produce an inflating homogeneous background state considering the potential in (\ref{lp4}). In this case, following the data from the map in Fig.~\ref{figureII}, initial conditions were chosen to produce a model expanding for 88 (approx.) e-folds, with $\phi_0=\sigma_0= 20\:M_{\mathrm{Pl}}$. The horizon and all the physical wavenumbers ($k_{\text{phys}}$) associated with each perturbation mode evolve according to the curves in plot in Fig.~\ref{fig:figureIV}.

To describe the evolution of the correlations between perturbations of the different scalar degrees of freedom, we rely on our knowledge of the correlators $\langle\Phi_A,\Phi_B\rangle$, $\langle\dot{\Phi}_A,\Phi_B\rangle$ and $\langle\dot{\Phi}_A,\dot{\Phi}_B\rangle$ to provide suitable (approximate) initial conditions at $(t=t_0)$.

Asymptotic consistency with the field amplitudes in Minkowski vacuum with a non-diagonal mass matrix is sufficient to determine these correlation matrices. Diagonalization of the mass matrix in the initial time surface provides estimated values of the correlations in $\langle\Phi_A,\Phi_B\rangle_{t_0}$:
\beq
&D_{AB}\equiv U_{AC}U^T_{DB}M^2_{CD},\nonumber\\
&\langle\Phi_A,\Phi_B\rangle_{t_0}=U^T_{AC}U_{DB}\left(\frac{a^{-3}}{2\hat{\omega}_{CD}}\right),
\eeq
where $\hat{\omega}_{CD}\equiv\sqrt{(k^2/a^2)\delta_{CD}+D_{CD}}$, $D_{AB}$ is the diagonal version of $M^2_{AB}$ and $U_{AB}$ is the orthonormal transformation matrix responsible of the diagonalization.

The classicality arguments stated in section \ref{sec:correlators} allow us to consider independent initial conditions for each component of the multiplet and its derivatives:
\beq
&\langle\Phi_A,\Phi_B\rangle_{t_0} = L_{AC}L^T_{CB},\nonumber\\ 
&\langle\dot{\Phi}_A,\Phi_B\rangle_{t_0} = \dot{L}_{AC}L^T_{CB}+L_{AC}L^T_{DB}\langle{\dot{\hat{\chi}}_C,\hat{\chi}_D}\rangle_{t_0} = 0,\nonumber\\
&\langle\dot{\Phi}_A,\dot{\Phi}_B\rangle_{t_0} = \dot{L}_{AC}\dot{L}^T_{CB}+L_{AC}L^T_{DB}\langle{\dot{\hat{\chi}}_C,\dot{\hat{\chi}}_D}\rangle_{t_0}\nonumber\\
&+\left(\dot{L}_{AC}L^T_{DB}-{L}_{AC}\dot{L}^T_{DB}\right)\langle{\dot{\hat{\chi}}_C,\hat{\chi}_D}\rangle_{t_0}.\label{initcond} 
\eeq 
Initial conditions for $\dot{L}_{AB}$ must be compatible with both the classicality $\langle\dot{\Phi}_A,\Phi_B\rangle_{t_0}=0$ and the antisymmetry of $\langle\dot{\hat{\chi}}_A,\hat{\chi}_B\rangle_{t_0}$, which leads us to $\dot{L}_{AB}=\langle\dot{\hat{\chi}}_A,\hat{\chi}_B\rangle_{t_0}=0$. Therefore, it is possible to use (\ref{initcond}) to obtain the corresponding initial state of every phase correlations. From (\ref{initcond}), we observe that the dependence of $\langle\dot{\hat{\chi}}_A,\dot{\hat{\chi}}_B\rangle_{t_0}$ on both $\langle\dot{\Phi}_A,\dot{\Phi}_B\rangle_{t_0}$ and $L^{-2}_{AB}(t_0)$ guarantees the supression of the fastest frequency scales at the time these reached their maximum values.

The scheme of evolution is also responsible for the efficiency of any spectral code. It represents the way in which modes are injected in the system and then evolved from subhorizon scales to a final instant of time outside the horizon.
We are interested of reducing the time in which mode evolution is irrelevant to the spectrum via time translational invariance. To achieve this, we release modes every five or ten e-folds from a constant energy scale $k_{\text{phys}}$ as in Fig.~\ref{fig:figureIV}, being aware of the initial conditions for the fluctuation modes and the background discussed previously. Each mode can be traced back into a physical wave mode leaving the horizon. This simple technique saves half of the evolution time necessary to evolve all the modes from any constant time surface.

\section{Results}\label{sec:results}

In order to evaluate the validity of the statements made so far, we produced two codes: the first one generates the correlation matrix by taking correlations of a random sample. In the second one, we solve the system given by (\ref{gaugefinal}), (\ref{eqmovdotdot}) and (\ref{eqmovdotchi}). In both scenarios, we followed the same initial conditions stated in the previous section. Also, we used the same solver in both cases, a symplectic eighth order accurate Gauss-Legendre integrator (see Ref.~\cite{Yoshida:1990zz} for more details).

In the first case, we used the Box-Muller algorithm to generate a set of four independent unit variance random vectors $(\hat{\chi}_A)$ satisfying (\ref{firstgauge}): two amplitudes and two \textit{velocities}. By using (\ref{ansatz}), we reproduce a certain number of field realizations (between 100-500) consistent with all the calculable two-point correlators -- i.e. $\langle\Phi_A,\Phi_B\rangle_{t_0}$, $\langle\dot{\Phi}_A,\Phi_B\rangle_{t_0}$ and $\langle\dot{\Phi}_A,\dot{\Phi}_B\rangle_{t_0}$-- at the surface of initial conditions. Each one of these realizations evolves in agreement with the equations of motion obtained from (\ref{pertact}).  We are aware that this is not the most efficient way to calculate correlations; nevertheless, it is worth to compare with the ``static'' version of the idea we are developing in this project.

\begin{figure}
\centering
\includegraphics[scale=0.4]{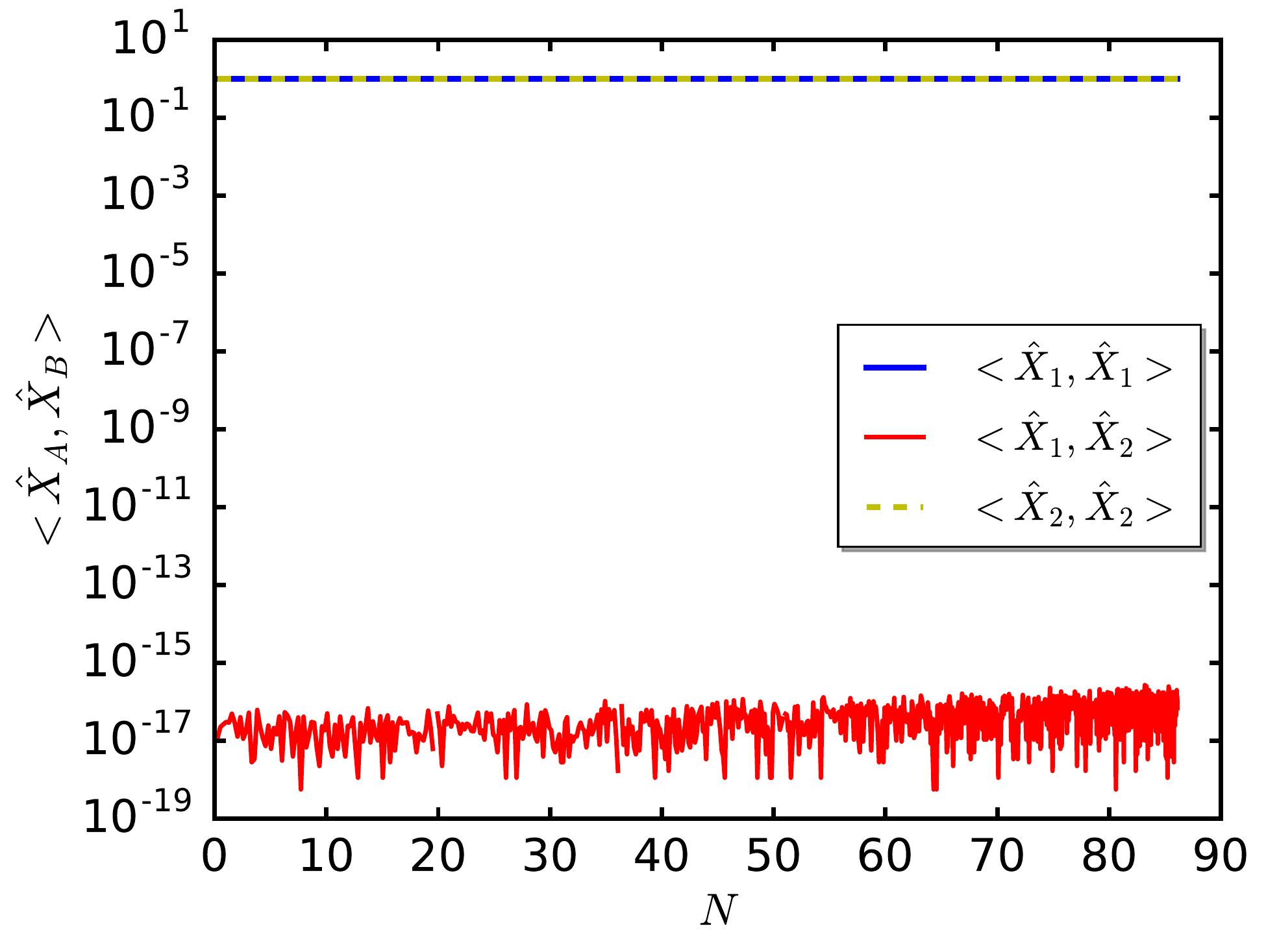}
\caption{\label{fig:gauge} The gauge choice $\langle\hat{\chi}_A,\hat{\chi}_B\rangle=\delta_{AB}$ holds at all times. Off-diagonal terms are numerically zero.} 
\end{figure}

\begin{figure}
\centering
\includegraphics[scale=0.4]{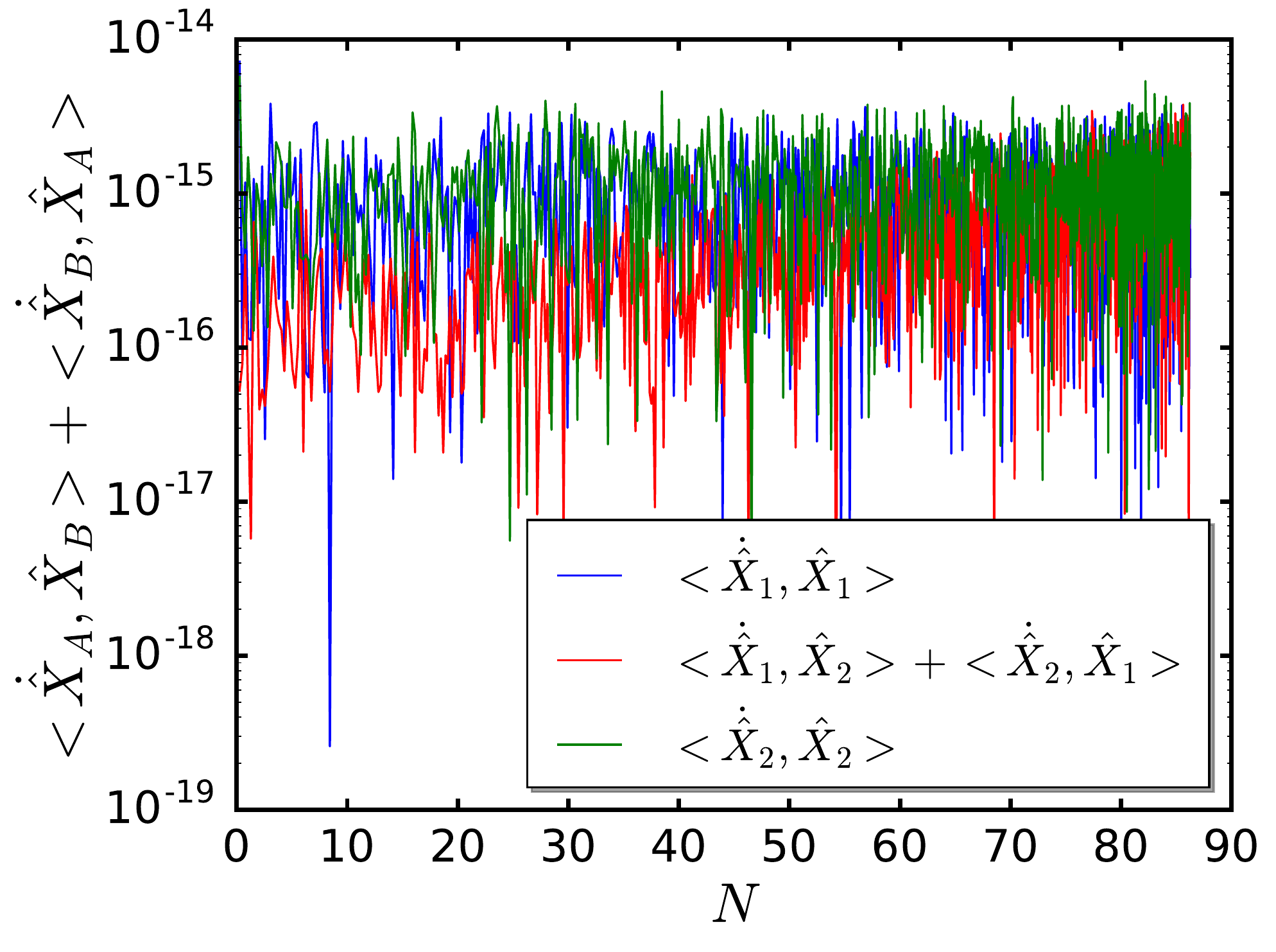}
\caption{\label{fig:gauge2} Antisymmetry of $\langle\dot{\hat{\chi}}_A,\hat{\chi}_B\rangle$.} 
\end{figure}

\begin{figure}
\centering
\includegraphics[scale=0.4]{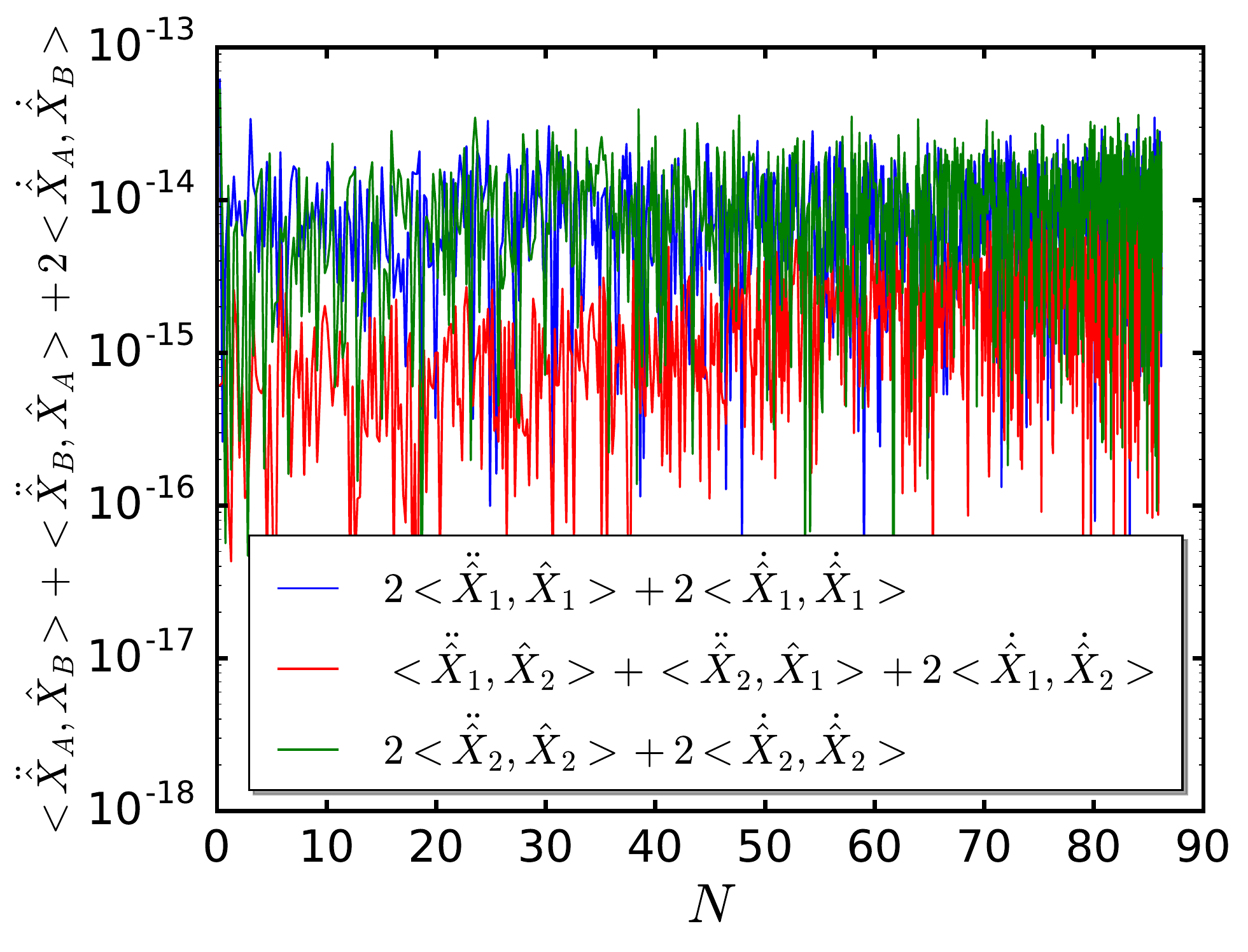}
\caption{\label{fig:gauge3} Verifying the final gauge condition in (\ref{thirdgauge}).} 
\end{figure}

By using the first code, we confirm by direct calculation the gauge fixing conditions provided on (\ref{firstgauge}), (\ref{secondgauge}) and (\ref{thirdgauge}). The dynamics of the system and the separation in fast and slow parameters relies entirely on these statements. In Fig.~\ref{fig:gauge}, we verified the first of these gauge choices.
After solving for $\Phi_1$ and $\Phi_2$, the inversion of (\ref{ansatz}) is used to confirm  (\ref{firstgauge}) as a valid gauge condition. The Cholesky decomposition of any positive definite correlation matrix is independent from the instant of time in which we choose to perform it. It is mainly supported by time translational invariance of the approximated Minkowski (or Bunch-Davies) vacuum state. In addition to this fact, we must also consider that this does not depend on the choice of any particular initial length scale. 

In Fig.~\ref{fig:gauge2} we check the numerical cancellation of the symmetric part of $\langle\dot{\hat{\chi}}_A,\hat{\chi}_B\rangle$.  
Phase cross-correlators $\langle\dot{\hat{\chi}}_1,\hat{\chi}_2\rangle$ and $\langle\dot{\hat{\chi}}_2,\hat{\chi}_1\rangle$ do not cancel separately. On the contrary, their strength is enhanced until the modes cross the horizon. It is possible to show that a Taylor expansion of $\langle\hat{\chi}_A,\hat{\chi}_B\rangle$ in powers of time, will only depend on $\langle\dot{\hat{\chi}}_A,\hat{\chi}_B\rangle$ and $\langle\dot{\hat{\chi}}_A,\dot{\hat{\chi}}_B\rangle$. The antisymmetric behaviour is consistent, again, with the notion of these correlators as rotation generators: the rotation transfers power from one mode to another.

In Fig.~\ref{fig:gauge3}, we tested the last dynamical gauge constraint in (\ref{thirdgauge}).
Here the combinations plotted are numerically zero. From this condition, we found the equations of motion of the ``square root'' factor $L_{AB}$.

Now we present the results of implementing the dynamical Cholesky scheme developed to get the expressions in (\ref{gaugefinal}), (\ref{eqmovdotdot}) and (\ref{eqmovdotchi}).

\begin{figure}
\centering
\includegraphics[scale=0.32]{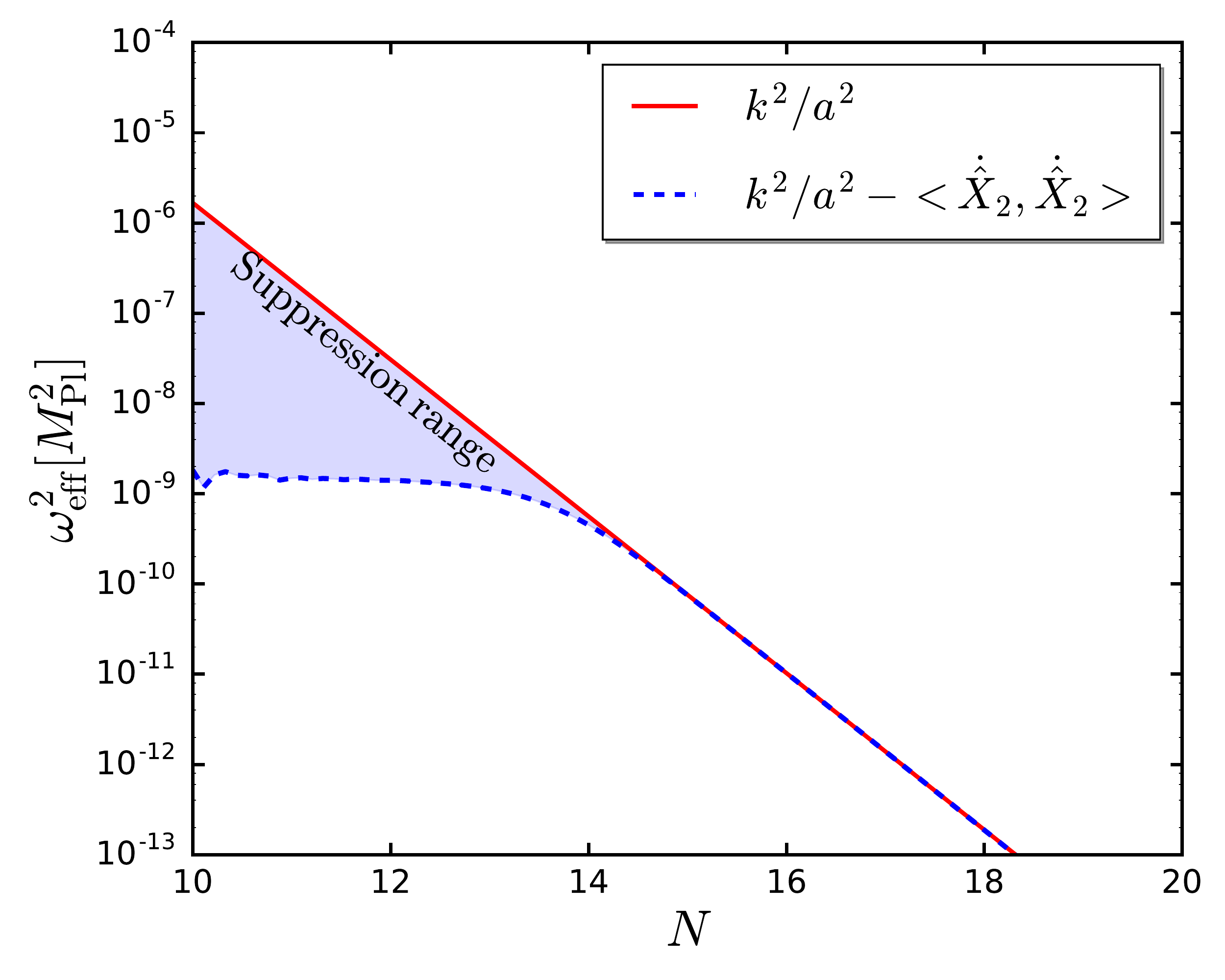}
\caption{\label{fig:omega} Suppressed oscillation scale in a mode injected at $k_1=e^{10}k_{\text{phys}}$. The effective oscillation frequency $(k^2/a^2)$ is reduced by three orders of magnitude. The effect due to $\langle\dot{\hat{\chi}}_1,\dot{\hat{\chi}}_1\rangle$ is the same.} 
\end{figure}

\begin{figure}
\centering
\includegraphics[scale=0.4]{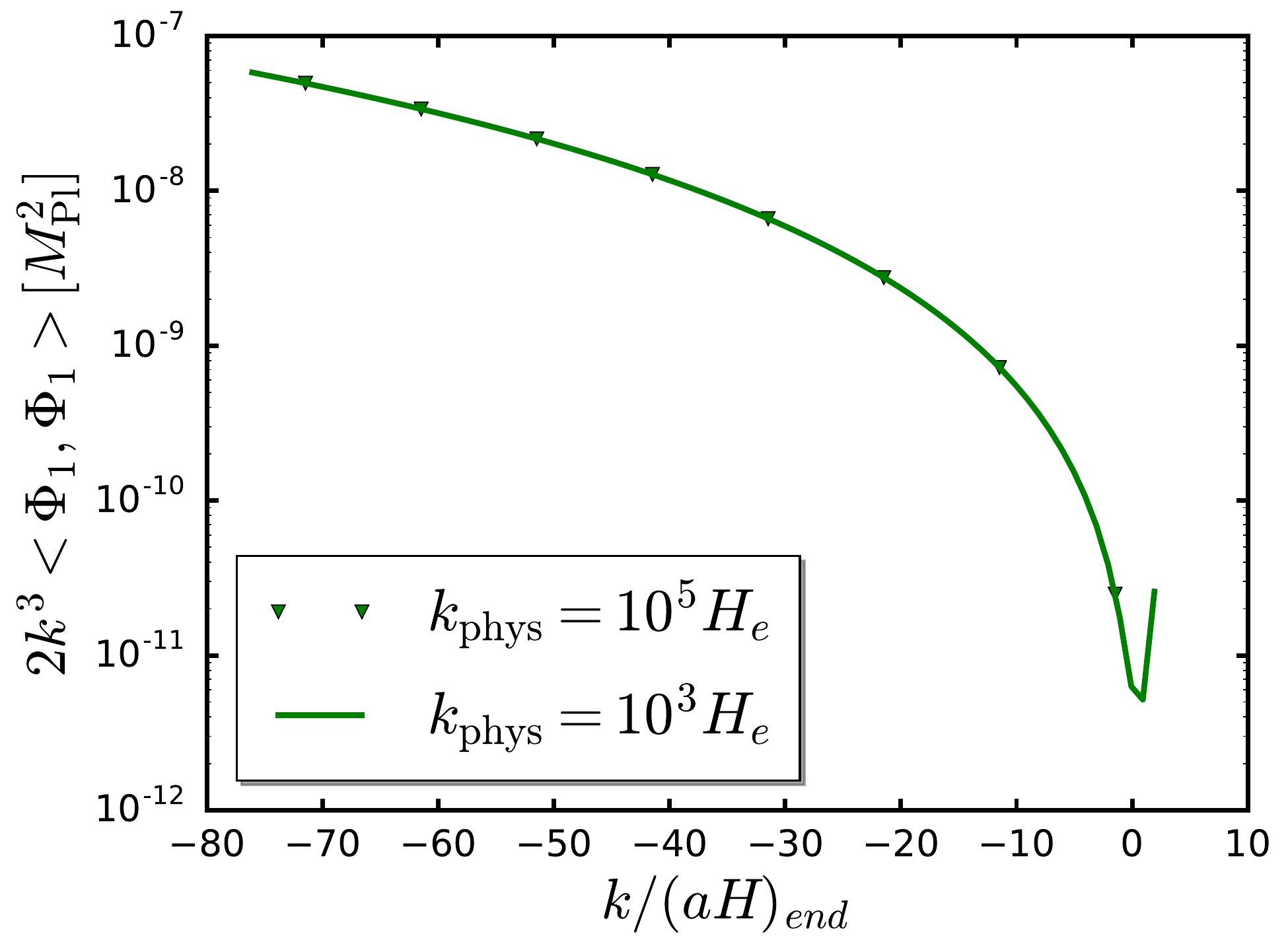}
\caption{\label{fig:invariance} $\langle\Phi_1,\Phi_1\rangle$ correlator obtained after injecting modes from two different physical length scales $k_{\text{phys}}=10^3 H_e$ and $k_{\text{phys}}=10^5H_e$. The spectrum remains unaltered.} 
\end{figure}

One of the objectives of this project is to separate the fast and slow degrees of freedom required to solve the spectrum of fluctuations. Such a separation is achieved by reducing the effective oscillation frecuency $\omega^2_{\mathrm{eff}}\sim k^2/a^2+M_{AB}^2$ by the introduction of $\langle\dot{\hat{\chi}}_A,\dot{\hat{\chi}}_B\rangle$ acting as counterterms in (\ref{gaugefinal}). In this particular case, even when the addition of the mass matrix alters the coupling between different field components, none of these magnitudes is enough to contribute significantly to $\omega^2_{\mathrm{eff}}$. In Fig.~\ref{fig:omega}, we can see an effective suppression of the higher frequency terms.
In the hypothetical case of a very large mass, the actual structure of the counterterms, provided by initial conditions (proportional to $\langle\dot{\hat{\Phi}}_A,\dot{\hat{\Phi}}_B\rangle_{t_0}^2\sim \omega_{\text{eff}}^2$) and equations of motion in (\ref{eqmovdotdot}), will affect the effective oscillating frequencies in exactly the same way. Therefore, due to this suppression, we can increase the time step required for resolving each Cholesky mode. Reducing the computational time remarkably when compared with other schemes.
    
In order to prevent convergence issues coming from any harmless inaccuracies in defining initial conditions (from either Minkowski or Bunch-Davies vacua), we evolve the modes for a couple of e-folds using smaller time steps. After a few iterations, the time step can be significantly increased. Consequently, modes can be injected from smaller length scales to increase the precision of the initial conditions for the correlators. 

In Fig.~\ref{fig:invariance}, we observe that there is no reason to expect any change in the shape of the spectrum if we decide to inject the modes from a different physical wavelength surface, even when this surface is deep inside subhorizon scales.

\begin{figure}[!ht]
\includegraphics[scale=0.4]{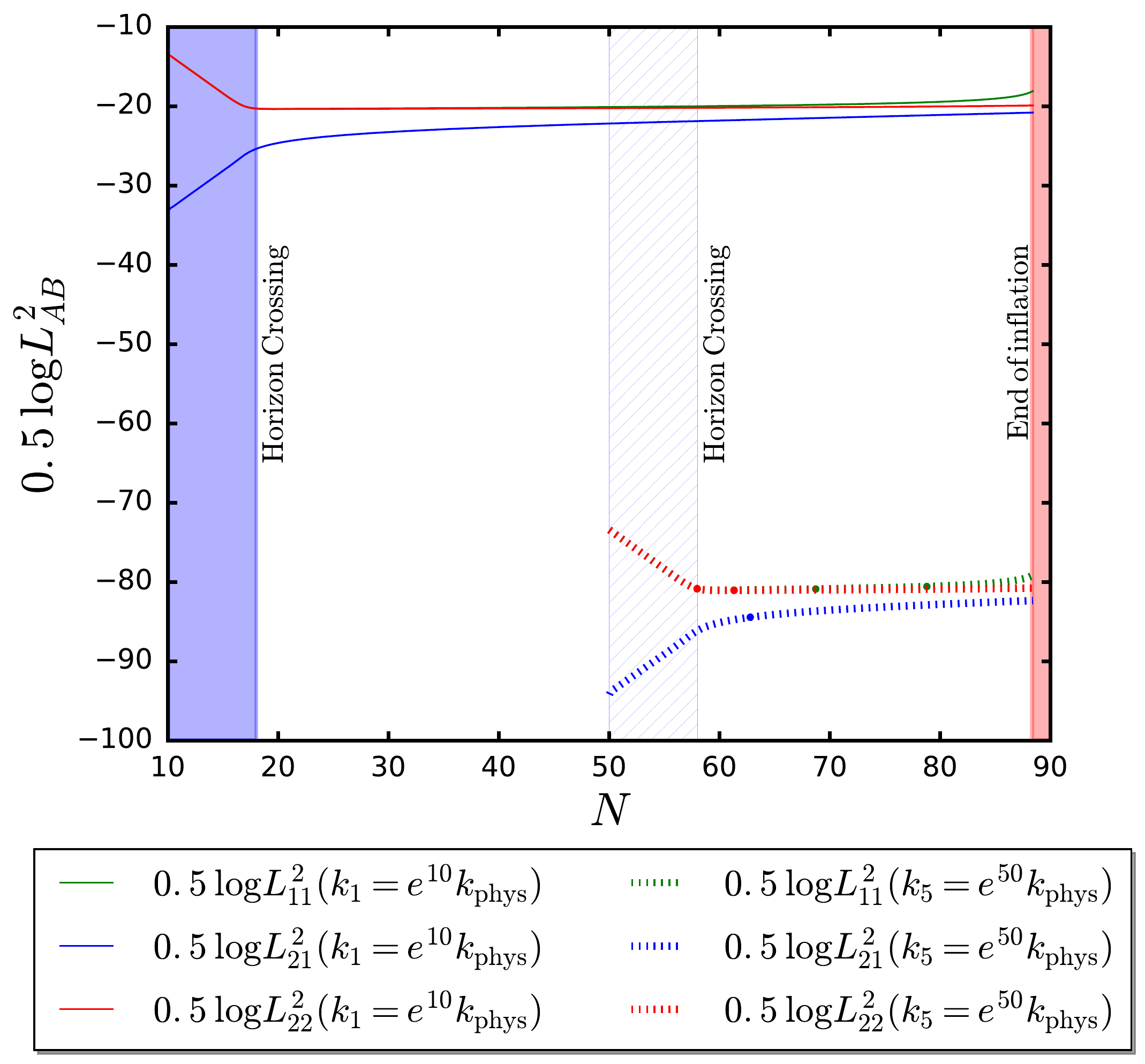}\label{amps0}
\vspace{0pt}
\caption{\label{fig:modeamps}  Evolution of two real Cholesky modes inserted at $N=10$ and $N=50$. We detect the suppression of oscillations before crossing the horizon (at $\ell=1/H$).}
\end{figure}

In Fig.~\ref{fig:modeamps}, we observe the evolution of two of the amplitude modes which constitute the spectrum injected from $k_{\text{phys}}=10^3\:H_e$, where $H_e$ is the Hubble parameter at the end of inflation. The injection scheme described in section \ref{sec:initial} is applied in order to collect the correlation amplitudes on a ``screen'' located at $N\approx 88.3$.  Notice the decay and the absence of oscillations before the horizon crossing.

In Appendix~\ref{App:AppendixB} we present our results for the two-point correlators as calculated by both of the approaches we used to find the spectrum. Results from the ``random sampling code'' are labeled as \textit{standard} and the outcome from the direct evolution of the Cholesky factors is labeled as \textit{gauged}. Both sets of results agree with a margin of relative uncertainty smaller than 0.1\%.  The speed enhancement in the \textit{gauged} routine allows us to insert more modes with negligible computational cost.

\begin{figure}
\includegraphics[scale=0.4]{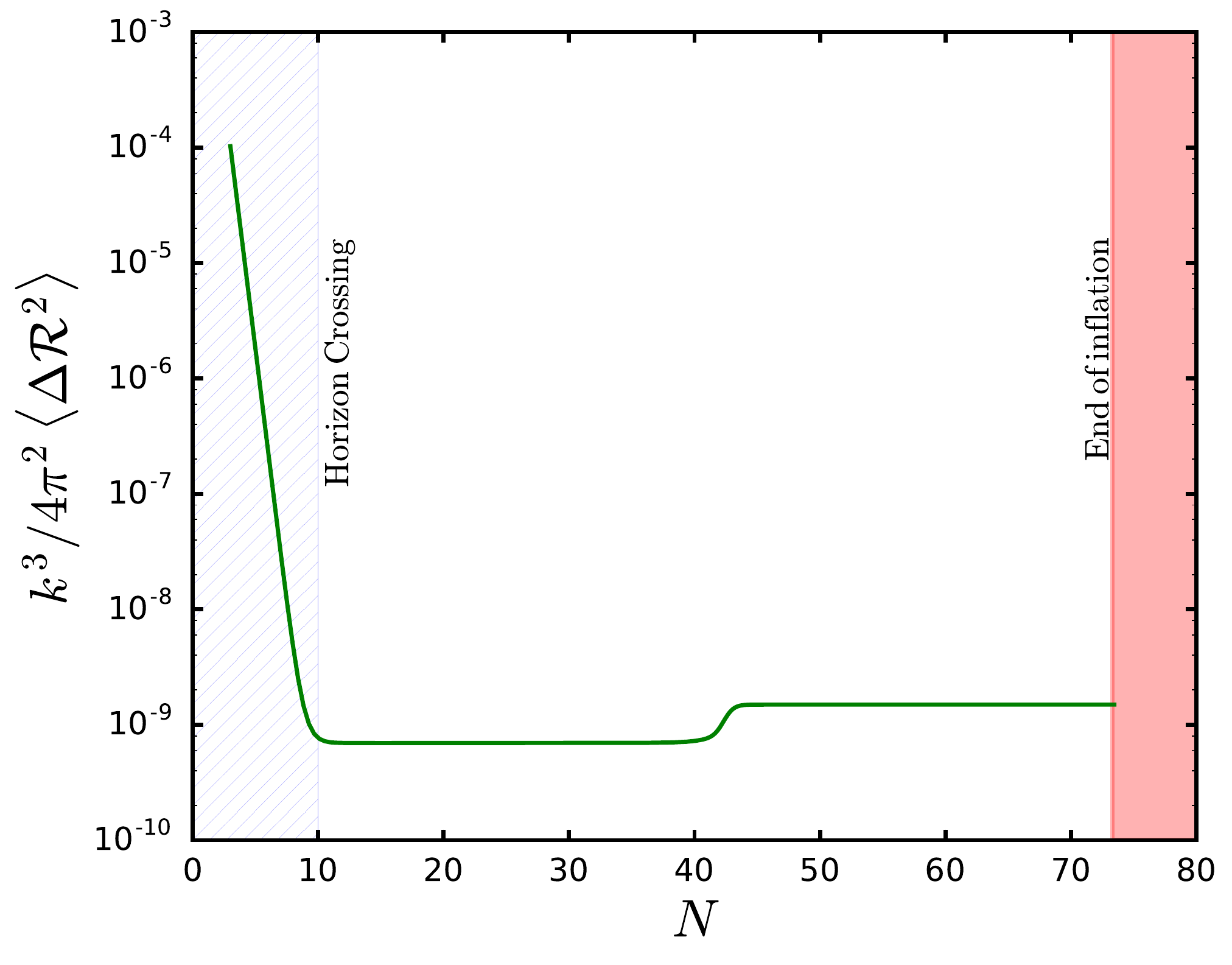}
\vspace{0pt}
\caption{\label{fig:curvature}  Evolution of $\langle\Delta\mathcal{R}^2\rangle$. Notice the change of the spectrum in super-horizon scales, it is now possible to inject the modes from smaller length scales and observe the earlier stages of evolution.}
\end{figure}

Using our method, it is also possible to calculate the comoving curvature perturbations, which are given by
\begin{equation}
\Delta\mathcal{R} = \frac{H}{\dot{\phi}_A\dot{\phi}_A}\left(\dot{\phi}_B\Phi_B\right),
\label{deltaR}
\end{equation} 
in the spatially flat gauge. Therefore, we find its corresponding power spectrum:
\begin{equation}
\langle\Delta\mathcal{R}^2\rangle_{t>t_0} = \frac{H^2}{\left(\dot{\phi}_A\dot{\phi}_A\right)^2}\left[\dot{\phi}_B\dot{\phi}_C\right]\langle\Phi_B,\Phi_C\rangle_{t>t_0},
\label{deltaR:correlator}
\end{equation} 
where $\langle\Phi_B,\Phi_C\rangle_{t>t_0}$ is now reconstructed at all instants of time by multiplying the Cholesky factors $L_{BD}$ as these evolve. As an additional result, we tested the production of primordial curvature fluctuations in the well-known double quadratic potential as proposed in Refs.~\cite{Price:2014xpa, Huston:2011fr, Lalak:2007vi, Avgoustidis:2011em, Frazer:2013zoa}
\begin{equation}
V(\phi,\chi)=\frac{m^2_{\phi}}{2}\phi^2+\frac{m^2_{\chi}}{2}\chi^2,
\label{m2m2}
\end{equation} 
at super-horizon scales. The shape of this potential is an elliptic paraboloid when we use the same parameters as in Ref.~\cite{Huston:2011fr}: $m_{\phi}=1.4\times10^{-6}M_{\mathrm{Pl}}$ and $m_{\chi}=7m_{\phi}$. Turning field trajectories are generated by the broken azimuthal symmetry of the potential in \eqref{m2m2}. We pick $\phi_0=\chi_0=12M_{\mathrm{Pl}}$ and $\dot{\phi}_0=\dot{\chi}_0=0$ as initial conditions for the background fields. In Fig.~\ref{fig:curvature} we use our decomposition method for a mode with $k\approx 1.3\times 10^5 H_\star$, where $H_\star$ is the Hubble parameter at the initial point of the field trajectory. We reproduce the spectrum of primordial curvature perturbations, which coincides exactly with the results obtained in Ref.~\cite{Huston:2011fr}. It is relevant to study turning trajectories in field space since these are the source of curvature perturbations even in more complicated potentials. 

\section{Discussions}
In this paper, we present a new method to separate fast and slow scales in the context of multifield models of inflation. We describe a scheme based on the Cholesky factorization of any positive definite correlation matrix.
 
As an outcome, we manage to find equations of motion for the ``square root'' of the correlation matrix in terms of slowly varying quantities. After reviewing the dynamical properties and contributions of the background fields, we specify a set of approximate initial conditions for the evolving system, and check that our results are consistent with straightforward averaging over all the realizations of the evolved random fields. We use the new code to calculate spectra of a few well-known models, and check that the shape of the spectrum depends on the choice of background trajectories.

This perspective based on dynamical Cholesky decomposition is significantly different from previous efforts in Refs.~\cite{Seery:2012vj, Price:2014xpa}. More recently, the Hamiltonian evolution of the field correlators $\langle\Phi_A,\Phi_B\rangle$ was considered in Ref.~\cite{Dias:2016rjq}, which is a transport scheme quite similar to ours in spirit. However, our method achieves significant computational gains by separating fast time scale of the sub-horizon mode evolution. Additionally, in our evolution scheme $\langle\hat{\chi}_A,\hat{\chi}_B\rangle=\delta_{AB}$ at every instant in time, which makes it convenient to generate properly-correlated random realizations of the fields for Monte-Carlo simulations.

This approach can be easily extendable to a diverse number of systems ruled by (almost) any perturbative manifestation arising from hyperbolic differential equations, which present fairly similar structures.  We will leave the non-Gaussian extension of this method in application to bispectrum computation for a future project.

\appendix
\section{Appendix: Decomposition scheme in a curved field space}\label{App:AppendixA}
As expressed in Ref.~\cite{Sasaki:1995aw}, we can consider a non-canonical version of (\ref{action})
\be
S = \int d^4x\sqrt{-g}\left(-\frac{1}{2}h_{AB}\partial_\mu\phi^A\partial^\mu\phi^B-V(\phi)\right).
\label{curvact}
\ee
From Refs.~\cite{Gong:2011uw,Elliston:2012ab}, we learned about the best way to consider the perturbative expansion of this action to get the equations of motion for the background
\beq
3M^2_{Pl}H^2&=&\frac{1}{2}\dot{\phi}_I\dot{\phi}^I+V(\phi)\label{curvedback}\\
\mathcal{D}_t\dot{\phi}_A+3H\dot{\phi}_A&=&\frac{\partial V}{\partial\phi_A},\nonumber
\eeq  
and a second order expansion of the action
\beq
&S_\Phi(k)={\displaystyle \frac{1}{2}\int} a^3\, d^4x\bigg[h^{AB}\mathcal{D}_t\Phi_A\mathcal{D}_t\Phi_B\nonumber\\
&-\Phi^A\left(\frac{k^2}{a^2}h_{AB}+\mathcal{M}_{AB}\right)\Phi^B\bigg],\label{curvedpert}
\eeq
where $\mathcal{D}_tX^A=\dot{X}^A+\Gamma^A_{CD}X^C\dot{\phi}^D$ and $\Gamma^A_{BC}=1/2h^{AD}\left(h_{BD,C}+h_{CD,B}-h_{BC,D}\right)$. Both expressions are fully covariant (and metric compatible) under translations along geodesics on the field space. The mass term $\mathcal{M}_{AB}$ is actively affected by the field space curvature
\be
\mathcal{M}_{AB}=\mathcal{D}^2_{AB}V-R_{DABC}\dot{\phi}^D\dot{\phi}^C-\frac{1}{a^3M^2_{Pl}}\mathcal{D}_t\left(\frac{a^3}{H}\dot{\phi}_A\dot{\phi}_B\right).\nonumber
\ee 

As we did before, we separate the field the field perturbations using $\Phi^A=L^{AB}\hat{\chi}_B$. Considering the case of a theory which produces equations of motion fully covariant under field space gauge transformations.  Hence, the form of the action only differs from (\ref{actchi}) by the use covariant derivatives instead of the total.

It is possible to promote our results (\ref{gaugefinal}), (\ref{eqmovdotdot}) and (\ref{eqmovdotchi}) after using the gauge fixing conditions for the correlators
\beq
&\langle\hat{\chi}_C,\hat{\chi}_D\rangle=h_{CD},\label{hinduced}\\
&\text{Sym}^{CD}_{AB} \langle\mathcal{D}_t\hat{\chi}_{C},\hat{\chi}_{D}\rangle=0\nonumber\\
&\text{Sym}^{CD}_{AB} \left[\langle\mathcal{D}^2_t\hat{\chi}_{C},\hat{\chi}_{D}\rangle+\langle\mathcal{D}_t\hat{\chi}_{C},\mathcal{D}_t\hat{\chi}_{C}\rangle\right]=0\nonumber
\eeq
and reshaping the form of the Cholesky amplitude matrix, for $(B<C)$ 
\beq
&\mathcal{A}_{BC}=-\left(L^{-1}\right)_{BE}\left(h^{EG}\mathcal{M}_{GD}\right)L^D_C-\nonumber\\
&(L^{-1})_{BE}\mathcal{D}_tL^{EF}\langle\mathcal{D}_t{\hat{\chi}}_F,\hat{\chi}_{C}\rangle+\langle\mathcal{D}_t\hat{\chi}_B,\mathcal{D}_t\hat{\chi}_C\rangle.\nonumber
\eeq 

We get:  
\begin{widetext}
\beq
&\mathcal{D}^2_tL^A_B+3H\mathcal{D}_tL^A_B+\left[\frac{k^2}{a^2}\delta^A_E+h^{AD}\mathcal{M}_{DE}\right]L^E_B+L^{AC}\mathcal{A}_{CB}+\mathcal{D}_tL^{AC}\langle\mathcal{D}_t\hat{\chi}_C\hat{\chi}_B\rangle-L^{AC}\langle\mathcal{D}_t\hat{\chi}_C\mathcal{D}_t\hat{\chi}_B\rangle=0,\nonumber\\
&\mathcal{D}_t\langle\mathcal{D}_t\hat{\chi}_A,\mathcal{D}_t\hat{\chi}_B\rangle=\text{Sym}^{CD}_{AB}\bigg[-3H\langle\mathcal{D}_t\hat{\chi}_C,\mathcal{D}_t\hat{\chi}_D\rangle-2L^{-1}_{CE}\dot{L}^{EF}\langle\mathcal{D}_t\hat{\chi}_F,\mathcal{D}_t\hat{\chi}_D\rangle-h^{EF}\mathcal{A}_{CE}\langle\mathcal{D}_t\hat{\chi}_F,\hat{\chi}_{D}\rangle+\nonumber\\
&L^{-1}_{CE}\mathcal{D}_tL^{EF}h^{GJ}\langle\mathcal{D}_t\hat{\chi}_F,\hat{\chi}_G\rangle\langle\hat{\chi}_{J},\mathcal{D}_t\hat{\chi}_D\rangle+h^{FG}\langle\mathcal{D}_t\hat{\chi}_C,\mathcal{D}_t\hat{\chi}_F\rangle\langle\mathcal{D}_t\hat{\chi}_G,\hat{\chi}_{D}\rangle\bigg],\nonumber\\
&\mathcal{D}_t\langle\mathcal{D}_t\hat{\chi}_A,\hat{\chi}_B\rangle= \mathcal{A}_{AB}-3H\langle\mathcal{D}_t\hat{\chi}_A,\hat{\chi}_B\rangle-\left(L^{-1}\right)_{AG}\mathcal{D}_t L^{GF}\langle\mathcal{D}_t\hat{\chi}_F,\hat{\chi}_B\rangle.
\eeq
\end{widetext}
Notice from (\ref{hinduced}) how the field metric is induced from the correlators of the fastest degrees of freedom in the system. The emergence of such a geometric structure appears as a result of averaging at each instant of time over the evolved gaussian random variables. 

\section{Appendix: Plotting two-point correlations}\label{App:AppendixB}

\begin{figure}
\begin{center}
\includegraphics[scale=0.3]{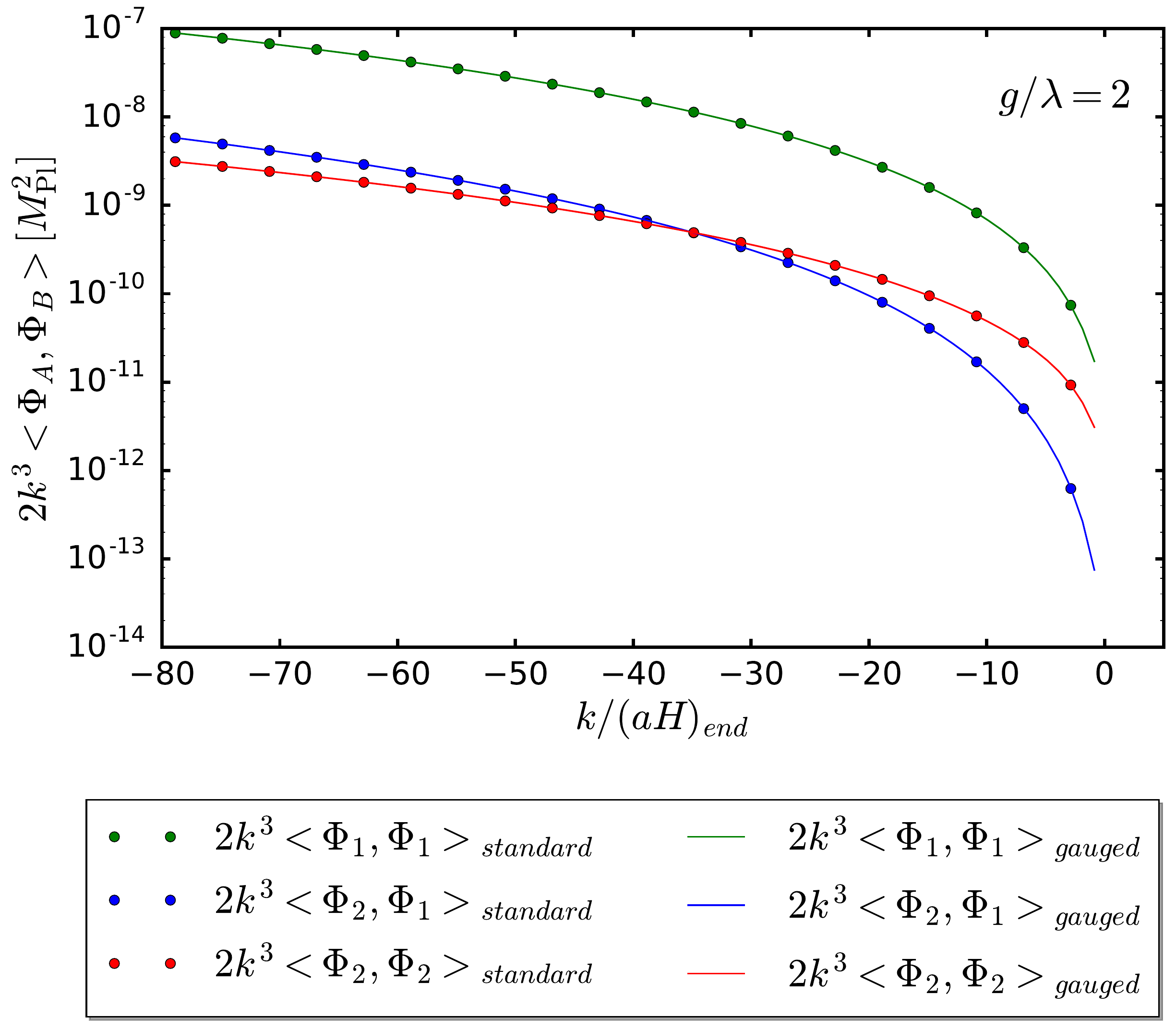}
\end{center}
\caption{\label{case1gl3} Scale invariant two-point correlators for the potential in (\ref{lp4}) for $g/\lambda=2$: The values of the two-point correlations obtained from evolving (\ref{gaugefinal}), (\ref{eqmovdotdot}) and (\ref{eqmovdotchi}) match with our results from averaging over 100-500 evolved field realizations.}
\bigskip

\begin{center}
\includegraphics[scale=0.25]{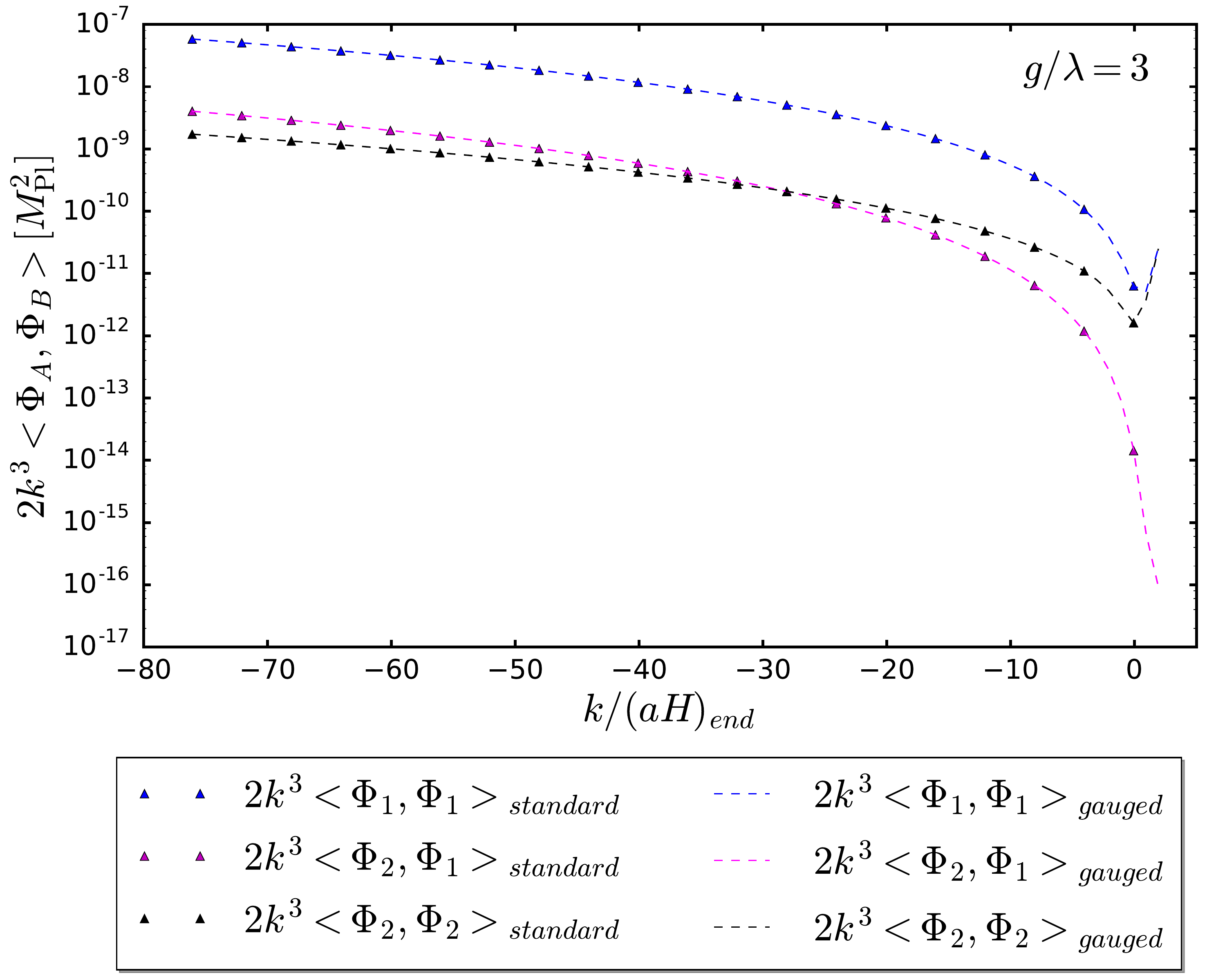}
\end{center}
\caption{\label{case1gl2} Two-point correlators for the case of $g/\lambda=3$.}
\end{figure}

In this this appendix, we show all the spectra calculated following the procedures aforementioned. In Fig.~\ref{case1gl3}, we provide the spectrum computed for the case of the specific background trajectory defined in section \ref{sec:initial} using the two methods suggested.

The spectrum represented in Fig.~\ref{case1gl2} corresponds to the case of the same initial conditions; now using a different ratio $g/\lambda=3$ between the coupling parameters in the potential (\ref{lp4}). 
Here $k/(aH)_{\text{end}}$ represents the normalization of comoving momentum with respect to the scale factor and the Hubble parameter measured at the end of inflation. In both cases, it is not possible to indicate which one of the field components dominates of the spectrum. This is due to our choice of the background field trajectories, in which the contribution of both fields is nearly the same. The only difference is the independent contribution of $\phi$ in (\ref{lp4}).

\begin{figure}
\begin{center}
\includegraphics[scale=0.31]{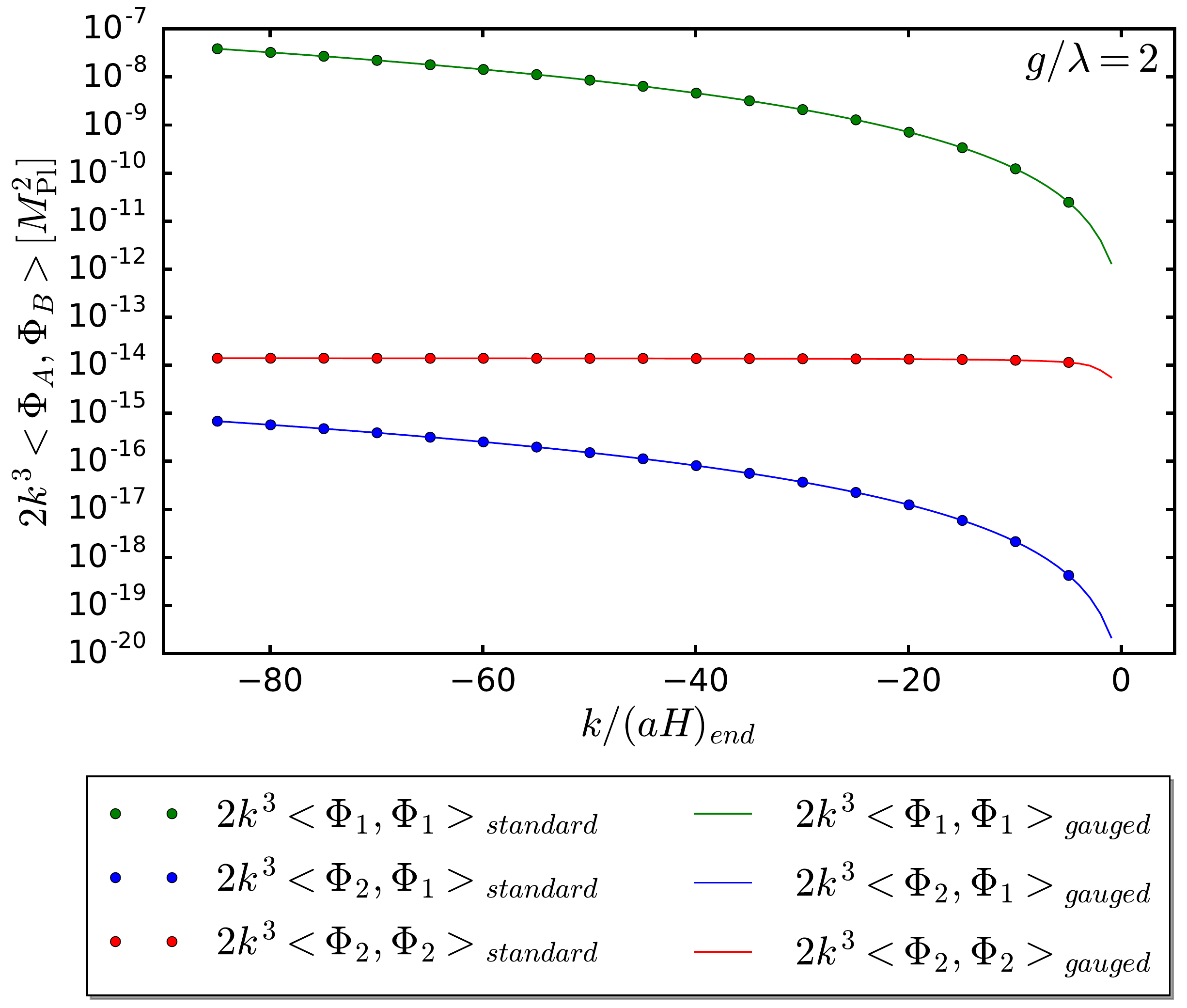}
\end{center}
\caption{\label{case2gl2} Two-point correlators for the case of $g/\lambda=2$, in the case of a dominant contribution from $\phi$.}
\bigskip

\begin{center}
\includegraphics[scale=0.25]{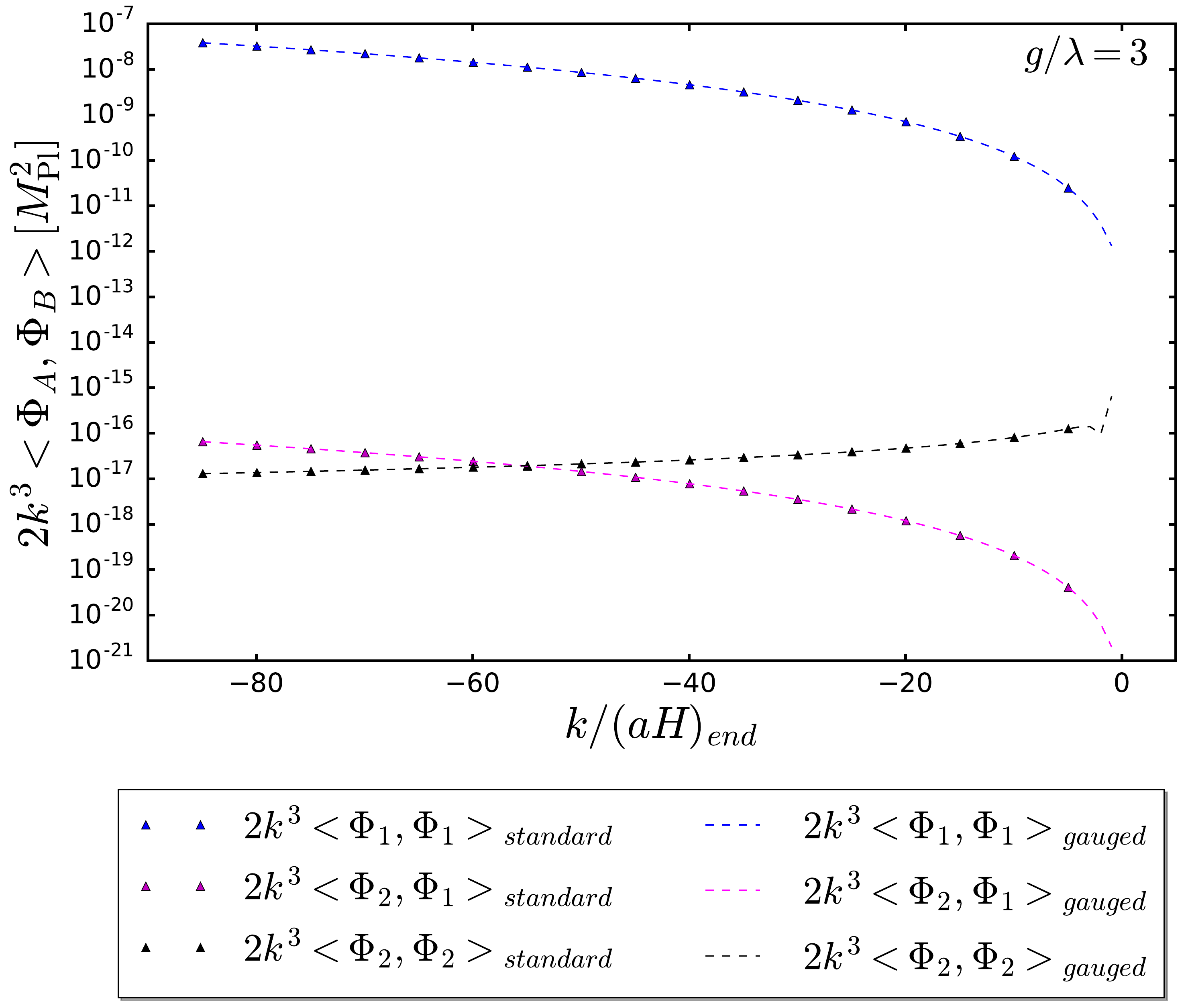}
\end{center}
\caption{\label{case2gl3} Two-point correlators for the case of $g/\lambda=3$, in the same case of dominance of $\phi$.}
\end{figure}

In Fig.~\ref{case2gl2} we show the spectrum for the case of transversal motion along the inflaton trajectory ($\sigma_0=0$) for $g/\lambda=2$. We achieve this by setting a dominant background contribution for $\phi$ in the same potential. 

Additionally, the case of $g/\lambda=3$ is represented in Fig.~\ref{case2gl3}.  Both spectra reflect the strength of the fluctuations of $\phi$ when compared to those from $\sigma$. Showing a suppression of the additional field and the cross correlators. The last results are fully consistent with the input required to generate the chaotic billiards pictures produced in Ref.~\cite{Bond:2009xx}.

It has been shown in Ref.~\cite{Gordon:2000hv} that the adiabatic and isocurvature components of the spectrum can be found via a change of basis.
Gram-Schmidt procedure provides an orthonormal basis of $\mathcal{N}$ vectors made of the background field components to project into the adiabatic/isocurvature basis just as in \eqref{deltaR}.

\newpage
\section*{ACKNOWLEDGEMENTS}
We would like to thank Jonathan Braden, Andrew DeBenedictis, Michael Desrochers, Zhiqi Huang, Malcolm Kennett, Claire Maulit, Justine Munich, Levon Pogosian, Sunny Vagnozzi and Alex Zucca for their valuable comments and discussions on the previous drafts of this paper. This project was partly funded by the Discovery Grants program of the Natural Sciences and Engineering Research Council of Canada.

\end{document}